\documentclass[11pt]{article}
\usepackage{amssymb,amsmath,amsthm}
\usepackage{graphicx}\DeclareGraphicsExtensions{.pdf, .jpg, .png , .bmp}
\usepackage{natbib}

\newtheorem{theorem}{Theorem}

\newtheorem{lemma}{Lemma}

\title{A Flexible Class of Non-separable Cross-Covariance Functions for Multivariate Space-Time Data}
\author{Marc Bourotte $^{a}$, Denis Allard $^{a}$ \& Emilio Porcu $^{b}$\\
\small{$^a$Biostatistique et Processus Spatiaux (BioSP),
  INRA, Avignon, France.}\\
\small{$^b$Departamento de Matem\'atica, Universidad T\'ecnica Federico Santa Mar\'ia, Valpara\'iso, Chile}
}

\begin{document}
\baselineskip24pt
\date{}

\maketitle

\noindent
Running title: Non-separable Multivariate Space-Time Cross-Covariance Functions.
\vskip2.5mm

\noindent
Corresponding author: D. Allard. 
INRA, UR546 BioSP, Site Agroparc\\
84914 Avignon, FRANCE\\
Tel: +33\,432722171; Fax: +33\,432722182\\
{\tt allard@avignon.inra.fr}
\vskip2.5mm

\section*{Abstract}

Multivariate space-time data are increasingly available in various scientific disciplines. When analyzing these data, 
one of the key issues is to describe the multivariate space-time dependencies. Under the Gaussian framework, one needs
to propose relevant models for multivariate space-time covariance functions, i.e. matrix-valued mappings with the 
additional requirement of non-negative definiteness.  
We propose a flexible parametric class of cross-covariance functions for multivariate space-time Gaussian random fields. 
Space-time components belong to the (univariate) Gneiting class of space-time covariance functions, with Mat\'ern or Cauchy 
covariance functions in the spatial margins. The smoothness 
and scale parameters can be different for each variable. We provide sufficient conditions for positive definiteness.
A simulation study shows that the parameters of this model can be efficiently estimated using weighted pairwise 
likelihood, which belongs to the class of composite likelihood methods. We then illustrate the model on a French dataset of weather variables.

\bigskip

\noindent{{\bf Keywords} Composite likelihood; Mat\'ern covariance; Multivariate Gaussian processes; Separability; Spatio-temporal processes; 
Spatio-temporal geostatistics.}

\section{Introduction}

Environmental and climate sciences provide an increasing amount of multivariate data indexed by space-time coordinates. 
For statisticians analyzing these data, one of the key issues is to model the space-time dependence structure, not only within each variable, 
but also between the variables. This requires models that allow for different range and smoothness parameters for each variable, 
and whose parameters can be accurately estimated. Let us recall some expository material that will be needed 
for the presentation of our results. 

\medskip

Consider a $p$-dimensional multivariate random field
$\textbf{Y}(\textbf{x})=\{Y_1(\textbf{x}),\dots,Y_p(\textbf{x})\}^\top$, where $Y_i(\textbf{x})$ represents the $i$-th variable, $i=1,\dots,p$, and
$\textbf{x} =(\textbf{s},t)\in D \times T \subset \mathbb{R}^{d+1}, d\geq 1$, where $\textbf{s}\in D \subset \mathbb{R}^d$ is a vector of spatial coordinates 
and $t \in T \subset \mathbb{R}$ is time. 
Let us further assume that $\textbf{Y}(\textbf{x})$ can be decomposed into the sum of a deterministic and a random component,
\begin{equation*}
\textbf{Y}(\textbf{x})=\boldsymbol{\mu}(\textbf{x})+\textbf{Z}(\textbf{x}),
\quad \textbf{x} \in D \times T,
\end{equation*}
where $\boldsymbol{\mu}(\cdot)$ is a trend function and $\textbf{Z}(\cdot)$  a zero mean multivariate 
Gaussian stationary process.
Under the assumptions of Gaussianiyt and stationarity, the process $\textbf{Z}(\cdot)$ is completely characterized by its 
matrix-valued covariance function $\textbf{C}(\textbf{h},u)=\left[C_{ij}(\textbf{h},u)\right]_{i,j=1}^p$, which
depends solely on the space-time lag, $\textbf{k}=(\textbf{h},u)\in\mathbb{R}^{d} \times \mathbb{R}$:
\begin{equation*}
{\rm Cov}\left\{Z_i(\textbf{s},t),Z_j(\textbf{s}+\textbf{h},t+u)\right\}=C_{ij}(\textbf{h},u), \quad i,j=1,\dots,p, \quad 
\textbf{s}, \textbf{s}+ \textbf{h} \in D,\quad t,t+u \in T.
\end{equation*}
Cross-covariance functions are not symmetric, but they are invariant with respect to the joint exchange of the variables and 
the sign of the lag $\textbf{k}$ : $C_{ij}(\textbf{k})=C_{ji}(-\textbf{k})$, for all $\textbf{k} \in \mathbb{R}^{d+1}$. 
Full symmetry is a more restrictive assumption for which the following
relationships are also verified: $C_{ij}(\textbf{k}) =C_{ij}(-\textbf{k}) =C_{ji}(\textbf{k})$. 
For a complete review on multivariate random fields modeling, with a particular focus on spatial cross-covariance functions, 
we refer the reader to  \citet{GentonKleiber2014} with the associated discussions.

\medskip

Our goal is to elaborate valid, flexible parametric classes of space-time matrix-valued covariance functions for 
$\textbf{Z}(\textbf{x})$, i.e. matrix-valued covariance functions that verify
the well-known requirement of non-negative definiteness: for any $n \in \mathbb{N}$, 
for any finite set of points $(\textbf{s}_1,t_1),...,(\textbf{s}_n,t_n)$ and for any vector $\boldsymbol{\lambda}\in\mathbb{R}^{np}$, we 
have $\boldsymbol{\lambda}^\top \boldsymbol{\Sigma} \boldsymbol{\lambda} \geq 0$, where $\boldsymbol{\Sigma}$ is a $np \times np$ matrix with 
$n  \times n$ block elements of $p \times p$ matrices
$\textbf{C}(\textbf{s}_\alpha-\textbf{s}_\beta,t_\alpha-t_\beta)$, with $\alpha,\beta=1,\dots,n$.

\medskip

Following \citet{GentonKleiber2014} and 
\citet{GelfandBanerjee2010}, a multivariate space-time covariance model is said to be separable when it is obtained through
the product of a $p \times p$ covariance matrix 
$\textbf{A}=[A_{ij}]_{i,j=1}^p$ and a valid univariate space-time correlation function  $\rho_{ST}(\cdot)$ on $D \times T$,
\begin{equation}
  \label{eq:space_time_sep}
\textbf{C}(\textbf{k})=\textbf{A}\rho_{ST}(\textbf{k}),\quad \textbf{k} \in D \times T.   
\end{equation}
When $\rho_{ST}(\cdot)$ is 
also space-time separable, the covariance matrix of  $\textbf{Z}$ reduces to 
\begin{equation}
\boldsymbol{\Sigma} = \textbf{A} \otimes \textbf{C}_S\otimes \textbf{C}_T,
\label{eq:full_sep}
\end{equation}
where $\otimes$ is the Kronecker product and $\textbf{C}_S,\textbf{C}_T$ are respectively spatial and 
temporal covariance matrices \citep{Cressie2011} associated to univariate spatial and temporal covariance functions.
Separability induces reduced number of parameters and faster computation of the inverse and of the determinant of the matrix $\boldsymbol{\Sigma}$.
Often, separability is an overly simplified  assumption for weather and climate data. Space-time separability is equivalent to conditional independence 
between $Z(\textbf{s},t)$ and $Z(\textbf{s}',t')$ given $Z(\textbf{s}',t)$ (or $Z(\textbf{s},t')$), since in this case we have
\begin{equation*}
{\rm Var}\left\{Z(\textbf{s},t)\right\} {\rm Cov}\left\{Z(\textbf{s},t),Z(\textbf{s}',t')\right\}=
{\rm Cov}\left\{Z(\textbf{s},t),Z(\textbf{s}',t)\right\}
{\rm Cov}\left\{Z(\textbf{s}',t),Z(\textbf{s}',t')\right\}, 
\end{equation*}
$(\textbf{s},t), (\textbf{s}',t') \in D \times T$. As a consequence, there is a proportional relationship between $C(\textbf{h},u)$ and $C(\textbf{h}',u)$ 
for two fixed spatial lags $\textbf{h},\textbf{h}'\in\mathbb{R}^d,u\in\mathbb{R}$, 
which implies that separable covariances cannot capture sophisticated interactions between space and time.
Moreover, in a multivariate framework, separability between variables and space-time variations implies that all variables 
are characterized by the same space-time correlation function. An approach based on the 
asymptotic distribution of the sample cross-covariance estimator to  test for separability and 
for full symmetry is proposed in \citet{Li2008}.  This allows the practitioner to select among 
the important dependence structures and to make the appropriate 
modeling choice. Often, separability must be rejected and models that do not separate space, time and variable index must be
defined.

\medskip

\begin{figure}
  \centering
  \includegraphics[width = 11cm, height = 5.5cm]{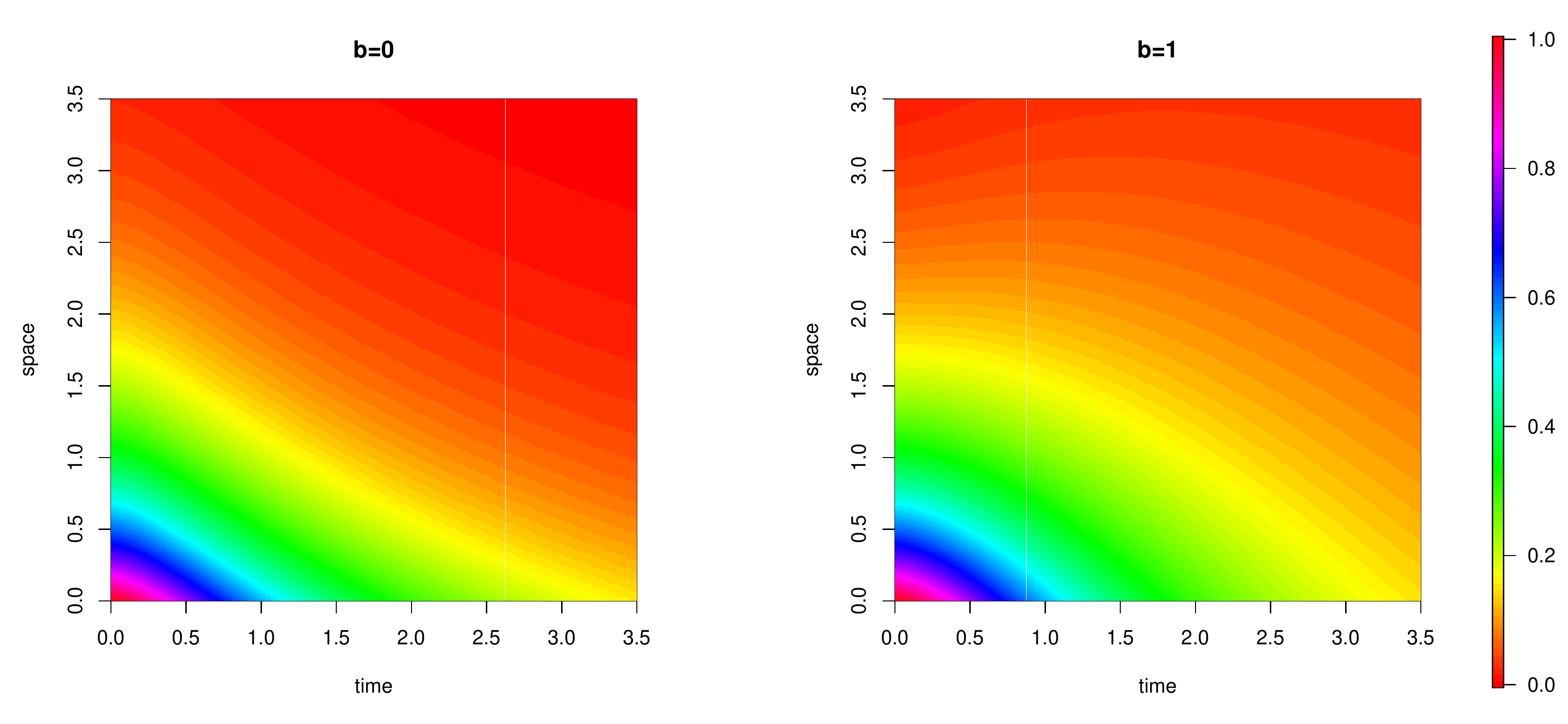}
  \caption{An example of a Gneiting space-time covariance function ${\cal G}(\textbf{h},u)=
(0.8|u|^{1.5}+1)^{-1}\exp\left\lbrace-\|\textbf{h}\|/(0.8|u|^{1.5}+1)^{-b/2}\right\rbrace$. The case $b=0$ corresponds to separability. \label{fig:Gneiting_example}}
\end{figure}

\citet{Gneiting2002} proposed a class of univariate fully symmetric space-time covariances, that has now become the standard 
class of models for univariate space-time Gaussian random fields in geostatistical applications; 
see also \citet{Gneiting2007} and references therein. The Gneiting class of space-time covariances
is defined as 
\begin{equation}
  \label{eq:Gneiting}
  {\cal G}(\textbf{h},u) = \frac{\sigma^2}{\psi(u^2 )^{d/2}}\ \varphi\left( \frac{\| \textbf{h}\|^2 }{\psi(u^2) }   \right), \quad (\textbf{h},u) \in \mathbb{R}^d \times \mathbb{R},
\end{equation}
where $\varphi: [0,\infty) \to \mathbb{R}$ is completely monotone on the positive real line, with $\varphi(0) < \infty$, $\psi$ is a positive function
whose derivative is completely monotone on the positive real line and $\sigma^2$ is a variance parameter. See \cite{PorcuZastavnyi2011} for more
relaxed necessary conditions and the complete characterization of this class. Figure \ref{fig:Gneiting_example} depicts an example of a Gneiting space-time covariance function.

\medskip

In spatial context, multivariate models for which space and variables are not separable
are easily built by combining separable ones. In the linear model of coregionalization (LMC, \citet{GoulardVoltz1992,Wackernagel2003}),
the matrix-valued covariance function is defined as 
a linear combination of separable models: ${\bf C}(\textbf{h}) = \sum_{k=1}^K \textbf{B}_k \rho_k(\textbf{h})$,
where $\rho_k(\textbf{h})$ are spatial covariance functions with $\rho_k(\textbf{0}) = 1$ and $\textbf{B}_k$ 
are positive definite matrices. Drawbacks of this construction have been discussed in \cite{Gneiting2010} and in \cite{Daley2014}.
\citet{DeIaco2013} proposed a space-time extension of this construction.
A major drawback of this approach is that the smoothness of any component of the multivariate random field
is that of the roughest process associated to $\rho_k(\cdot),\ k=1,\dots,K$. 
\citet{Apanasovich2012} introduced a valid parametric family of cross-covariance functions for multivariate spatial random fields where each component has a covariance function from a Mat\'ern class. 

In space-time context, \citet{Apanasovich2010} proposed valid non-separable matrix-valued space-time covariance functions through additional 
latent dimensions, one for each of the $p$ variables, 
that represent the variables to be modeled. These dimensions are then used in a similar way as that of time within the Gneiting class
of covariance functions. Climate and weather variables need flexible space-time models able to capture different 
regularity properties from one variable to another. Under the framework proposed in  \citet{Apanasovich2010}, this can be obtained 
through a LMC-type construction, at the cost of a significant increase of the number of parameters.
In order to keep a reasonably low number of parameters, we argue in favor of a different approach.

\medskip

The Mat\'ern family \citep{Matern1986} of covariance functions  has become extremely popular in geostatistical applications. Its expression is
\begin{equation*}
{\cal M}(\textbf{h};r,\nu)=\frac{\sigma^2 2^{1-\nu}}{\Gamma(\nu)}\left(r\|\textbf{h}\|\right)^\nu \mathcal{K}_\nu\left(r\|\textbf{h}\|\right),\ \ \ \textbf{h}\in\mathbb{R}^d,
\end{equation*}
where $r>0$ is a scale parameter ($1/r$ is often called the range) and  $\mathcal{K}_\nu$ denotes the modified Bessel function of the second 
kind of order $\nu$ \citep{NIST:DLMF}. The smoothness parameter $\nu>0$ is directly related to the regularity
of the underlying random field. Specifically, a random field $Y(\textbf{x})$ with Mat\'ern covariance is $m$ times mean square differentiable if and only if $\nu \geq m$. 
As $\nu \to \infty$, the random field is infinitely mean square differentiable and its associated covariance function tends to the so-called
Gaussian covariance function $C(\textbf{h}) = \exp( -r^2 \|\textbf{h}\|^2),\ \textbf{h} \in \mathbb{R}$. 
Small values of $\nu$ yield rougher random fields; in particular 
$\nu = 1/2$ corresponds to the exponential covariance function $C(\textbf{h}) = \exp( -r \|\textbf{h}\|)$ that is mean square continuous but not mean
square differentiable at the origin. When $\nu = k + 1/2$ and  $k$ is a positive integer, the Mat\'ern model reduces to the product of the negative 
exponential with a polynomial of degree $k$.

A multivariate version of the Mat\'ern family has been proposed in \citet{Gneiting2010} and further extended in 
\citet{Apanasovich2012}. These constructions have the same general structure with
\begin{equation}
C_{ij}(\textbf{h})=\sigma_i\sigma_j\rho_{ij} {\cal M}(\textbf{h};r_{ij},\nu_{ij})
,\ \ \ \textbf{h}\in\mathbb{R}^d,
\label{eq:MultiMatern}
\end{equation} 
with different restrictions on the parameters $\{r_{ij},\nu_{ij},\rho_{ij}\}_{i,j = 1,\dots,p}$ to ensure the validity of the 
matrix-valued covariance function $\textbf{C}(\textbf{h})=\left[C_{ij}(\textbf{h})\right]_{i,j=1}^p$. In this model, each marginal component 
of $\textbf{Z}$, i.e. each covariance function $C_{ii}(\textbf{h})$,
has a different smoothness parameter,
while allowing some cross correlations between the variables. Sufficient conditions have been obtained using scalar 
mixtures \citep{Gneiting2010, Schlather2010,PorcuZastavnyi2011} 
while necessary and sufficient conditions are provided in \citet{Gneiting2010} for $p=2$ only.

\medskip

As an alternative to the Mat\'ern class, it is also possible to build a multivariate version of the generalized 
Cauchy family, having expression 
\begin{equation*}
\mathcal{C}(\textbf{h};r,\beta,\lambda)=\left(1+r\|\textbf{h}\|^\lambda\right)^{-\beta/\lambda}
,\ \ \ \textbf{h}\in\mathbb{R}^d,
\end{equation*}
with $0<\lambda\leq 2$ and $\beta,r>0$. In \citet{GneitingSchlather2004} it is shown that $\lambda$ characterizes the roughness of the associated random
field. Higher values of $\lambda$ yielding smoother fields, while $\beta$ parametrizes the dependence at large distances.
Another parameterization, which will be used in the rest of this work, is obtained 
by replacing $\beta / \lambda$ by $\nu>0$.
The isotropic Mat\'ern and Cauchy covariance functions are related to the 
completely monotone functions of Table 1 in \citet{Gneiting2002} by the equation $C(\textbf{h})=\varphi(\|\textbf{h}\|^2)$ \citep{Schoenberg1938}.

\medskip

In this article we propose a  class of non-separable multivariate space-time covariance models for any number of variables. 
The general structure of this class is 
${\bf C}(\textbf{h},u) = [C_{ij}(\textbf{h},u)]_{i,j=1}^p$ with
\begin{equation*}
C_{ij}(\textbf{h},u)=\frac{\sigma_i\sigma_j}{\psi(u^2)^{d/2}}\rho_{ij}\varphi_{ij}\left(\frac{\|\textbf{h}\|^2}{\psi(u^2)}\right)
 ,\ \ 1 \le i,j \leq p, \ \ (\textbf{h},u)\in\mathbb{R}^{d}\times\mathbb{R}.
\end{equation*} 
It is an extension of the Gneiting class $\cal G$ to the multivariate case, where the completely monotone functions $\varphi_{ij}$ have different 
parameters for each variable.

In particular, we offer sufficient conditions for two cases: $\varphi_{ij}(t) = {\cal M}(t^2;r_{ij},\nu_{ij})$ and 
$\varphi_{ij}(t) = {\cal C}(t^2;r_{ij},\nu_{ij},\lambda)$.  The parameterization is inspired by
the multivariate Mat\'ern structure in \citet{Gneiting2010} and \citet{Apanasovich2012} for spatial Gaussian random fields.
For both classes, each variable has its own range and own degree of smoothness, while allowing for cross-correlation.  
The remainder of the article is organized as follows. Section 2 presents the new multivariate space-time classes. 
Sufficient conditions that provide valid matrix-valued covariance functions are established and proved in the Appendix.
An estimation technique, based on pairwise composite likelihood, is then presented in Section 3. Section 4 is devoted to a 
simulation study in order to validate the estimation procedure. 
In Section 5, the Mat\'ern model is applied to a multivariate space-time dataset of weather variables in 
France. It is shown that the proposed model offers substantial 
improvements to separable models and also to more parsimonious models with equal range and smoothness parameters for all variables.

\section{A class of non-separable space-time cross-covariance functions}

The mixture approach described in \citet{Gneiting2010} to build Mat\'ern matrix-valued covariance functions for multivariate random fields
is based on the stability properties of positive definite functions \citep[Proposition 3.1]{Reisert2007}.
It uses the scale mixture representation of the Mat\'ern covariance function \citep[Eq. 3.471.9]{Gradshteyn2007}. We extend it to multivariate
space-time random fields to build valid matrix-valued covariance functions in $\mathbb{R}^d \times \mathbb{R}$. It can also be adapted to other covariance
functions allowing mixture representations, such as the Cauchy model. Let us first recall the mixture representation in the space-time
framework.

\medskip

\begin{lemma}
Let the matrix-valued function $\textrm{\bf m}(\xi): \mathcal{Q}\subset\mathbb{R} \to \mathbb{R}^p \times \mathbb{R}^p$ be
symmetric and non-negative definite for all $\xi\in\mathcal{Q}$.  Let $C_\xi:\mathbb{R}^{d}\times\mathbb{R}\rightarrow\mathbb{R}$ 
be a univariate space-time covariance function for any fixed $\xi\in\mathcal{Q}$. 
Suppose furthermore 
that for all $i,j=1,\dots,p$ and fixed $(\textbf{h},u) \in \mathbb{R}^d \times \mathbb{R}$
the product $m_{ij}(\cdot)C_\xi(\cdot)$ is integrable with respect to a positive measure $F$ on $\mathcal{Q}$. 
Then,  the matrix-valued function $\textbf{C}(\textbf{h},u)=\left[C_{ij}(\textbf{h},u)\right]_{i,j=1}^p$
defined through
\begin{equation}
C_{ij}(\textbf{h},u)=\int_{\mathcal{Q}} m_{ij}(\xi) C_\xi(\textbf{h},u)F( {\rm d}\xi),\ \ 1\leq i,j \leq p,
\label{eq:eq_proof}
\end{equation}
is a valid $p$-variate matrix-valued covariance function on $\mathbb{R}^{d}\times\mathbb{R}$. 
\end{lemma}
This lemma is a direct application  of Theorem 1 in \cite{PorcuZastavnyi2011}. With appropriate choices for $C_\xi(\cdot)$ and $\textbf{m}(\cdot)$
the following results can be established.

\begin{theorem}
\label{theo:GM}
Let $\psi(t),t\geq 0,$ be a positive function with a completely monotone derivative. The multivariate  Gneiting-Mat\'ern space-time model denoted
$\textbf{C}^{\cal M}=\left[ C^{\cal M}_{ij}(\cdot,\cdot) \right]_{i,j=1}^p$, with
\begin{equation}
C^{\cal M}_{ij}(\textbf{h},u)=\frac{\sigma_i\sigma_j}{\psi(u^2)^{d/2}}\rho_{ij}{\cal M}\left(\frac{\textbf{h}}{\psi(u^2)^{1/2}}; 
r_{ij},\nu_{ij}\right),\ \ \ (\textbf{h},u)\in\mathbb{R}^d\times\mathbb{R},
\label{eq:new_model_Matern}
\end{equation}
defines a valid matrix-valued covariance function if, for all $i,j = 1,\dots,p$, 
\begin{eqnarray*}
r_{ij}& = &\{(r_i^2+r_j^2)/2\}^{1/2},\\
\nu_{ij}& = &(\nu_i+\nu_j)/2,\\
\rho_{ij}& = &\beta_{ij}\frac{\Gamma(\nu_{ij})}{\Gamma(\nu_i)^{1/2}\Gamma(\nu_j)^{1/2}}\frac{r_i^{\nu_i}r_j^{\nu_j}}{r_{ij}^{2\nu_{ij}}},
\end{eqnarray*}  
with $r_i, \nu_i >0$ for all $i =1,\dots,p$, and where $\boldsymbol{\beta} = \left[\beta_{ij}\right]_{i,j=1}^p$ is a correlation matrix.
\end{theorem}

\begin{theorem}
\label{theo:GC}
Let $\psi(t),t\geq 0,$ be a positive function with a completely monotone derivative. The multivariate Gneiting-Cauchy space-time model denoted
$\textbf{C}^{\cal C}=\left[ C^{\cal C}_{ij}(\cdot,\cdot) \right]_{i,j=1}^p$, with
\begin{equation}
C^{\cal C}_{ij}(\textbf{h},u)=\frac{\sigma_i\sigma_j}{\psi(u^2)^{d/2}}\rho_{ij}\mathcal{C}\left(\frac{\textbf{h}}{\psi(u^2)^{1/2}};
r_{ij},\nu_{ij},\lambda\right),\ \ \ (\textbf{h},u)\in\mathbb{R}^d\times\mathbb{R},
\label{eq:new_model_Cauchy}
\end{equation}
is a valid matrix-valued covariance function if, for all $i,j = 1,\dots,p$, 
\begin{eqnarray*}
r_{ij}& = &\{(r_i^{-1}+r_j^{-1})/2\}^{-1},\\
\nu_{ij}& = &(\nu_i+\nu_j)/2,\\
\rho_{ij}& = &\beta_{ij}\frac{\Gamma(\nu_{ij})}{\Gamma(\nu_i)^{1/2}\Gamma(\nu_j)^{1/2}}\frac{r_{ij}^{\nu_{ij}}}{(r_i^{\nu_i}r_j^{\nu_j})^{1/2}},
\end{eqnarray*}  
with $r_i, \nu_i >0$ for all $i =1,\dots,p$ and $0<\lambda\leq 2$, and where $\boldsymbol{\beta} = \left[\beta_{ij}\right]_{i,j=1}^p$ is a correlation matrix.
\end{theorem}

\begin{figure}
  \centering
  \includegraphics[width = 14cm, height = 9cm]{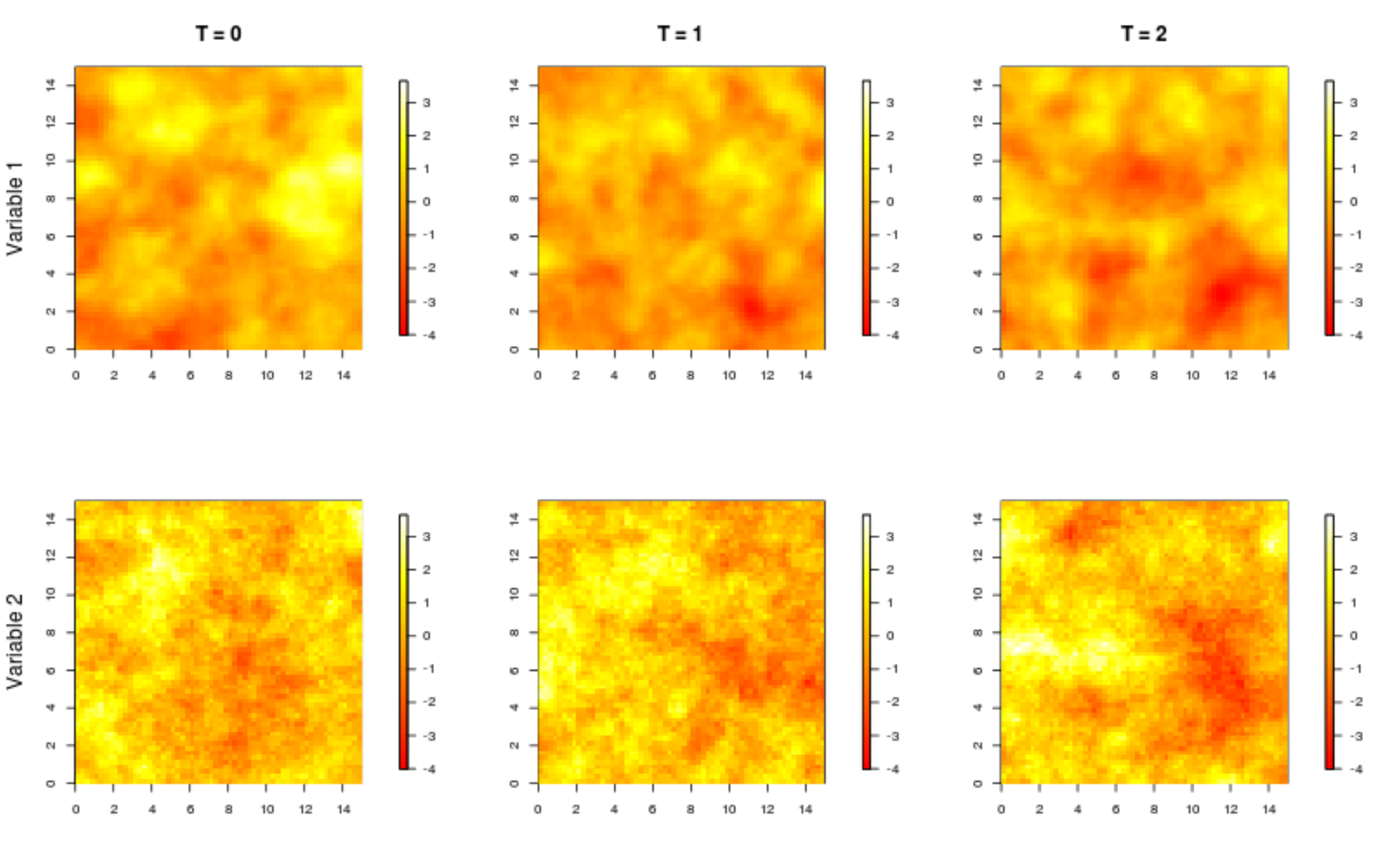}
  \caption{Bivariate Gneiting-Mat\'ern space-time Gaussian random field at 3 consecutive instants. 
Top row: smoother component  ($\nu_1=1.5$ and $r_1=1$); bottom row: rougher component ($\nu_2=0.5$ and $r_2=0.5$). 
The co-located correlation parameter  $\rho_{12}$ is set to 0.5.\label{fig:ex_simu}}
\end{figure}

The proofs are deferred to the Appendix.   
Each marginal space-time covariance function $C_{ii}^{\cal F}(\cdot,\cdot)$, $i=1,\dots,p$, ${\cal F} \in\{{\cal M}, {\cal C}\}$,  belongs to 
the Gneiting class with different regularity parameters $\nu_i$, $i=1,\dots,p$ for its associated 
Mat\'ern, respectively Cauchy, spatial covariance function.  
The Gneiting-Mat\'ern model is a space-time generalization of one 
particular case of the multivariate spatial ``flexible model'' presented in \citet{Apanasovich2012}. 
Figure \ref{fig:ex_simu} shows a  realization of 
a bivariate Gaussian field with Gneiting-Mat\'ern space-time covariance
at instants $t\in\{0,1,2\}$. In this example, the first component is smooth, with $\nu_1=1.5$ and 
scale parameter $r_1=1$, whereas the second component is rough, with $\nu_2=0.5$ and $r_2=0.5$. 
The co-located correlation parameter is $\rho_{12}=0.5$.

\medskip

Positive functions with completely monotone derivative, sometimes referred to as Bernstein functions \citep{Porcu2011}, 
are related to isotropic variograms by the following relationship: if a function $\psi(x),\ x \geq 0,$ belongs to the class of Bernstein functions,
the function  $\gamma( \textbf{h} )=\psi(\| \textbf{h}\|^2) -c,  \textbf{h} \in\mathbb{R}^d$ is a valid isotropic variogram for some constant $c$.  
For the rest of this work, we will consider the parametric form advocated in \citet{Gneiting2002} and \citet{Schlather2010}
\begin{equation*}
\psi(x)=(\alpha x^a+1)^b,\ \ \ x\geq 0,
\end{equation*} 
with $\alpha>0,0 < a\leq 1,0\leq b \leq 1$. With this choice (and a slight abuse of notation, since the Mat\'ern covariance 
does not have a parameter $\lambda$), the above models become
\begin{equation}
C_{ij}(\textbf{h},u)=\frac{\sigma_i\sigma_j}{(\alpha |u|^{2a} +1)^{bd/2}}\rho_{ij}{\cal F}\left(\frac{\textbf{h}}{(\alpha |u|^{2a} +1)^{b/2}};
r_{ij},\nu_{ij},\lambda,\right),\ \ \ (\textbf{h},u)\in\mathbb{R}^d\times\mathbb{R},
\label{eq:new_model}
\end{equation}
with ${\cal F} \in \{ {\cal M}, {\cal C}\}$ and the parameter restrictions stated in Theorems \ref{theo:GM} and \ref{theo:GC}. Note that
in the parameterization (\ref{eq:new_model}), the space-time
non-separability parameter $b$ acts both on the spatial covariance $\cal F$ and on the temporal one. A major drawback of this parameterization 
is that in case 
of space-time separability, i.e. when
$b=0$, the temporal covariance is equal to 1 for all $u \in \mathbb{R}$. Following  \citet{Gneiting2002}  a reparameterization is useful. Multiplying 
(\ref{eq:new_model}) by the temporal covariance function $(\alpha |u|^{2a} +1)^{-\delta/2}$, $u \in \mathbb{R}$, with $\delta >0$
and replacing the exponent $\delta + bd/2$ by $\tau$ 
leads to the parametric family, which will be used in the rest of this work
\begin{equation}
C_{ij}(\textbf{h},u)=\frac{\sigma_i\sigma_j}{(\alpha |u|^{2a} +1)^{\tau}}\rho_{ij}{\cal F}\left(\frac{\textbf{h}}{(\alpha |u|^{2a} +1)^{b/2}};
r_{ij},\nu_{ij},\lambda,\right),\ \ \ (\textbf{h},u)\in\mathbb{R}^d\times\mathbb{R},
\label{eq:new_model_final}
\end{equation}
with $\tau \geq bd/2$.
This parameterization provides independent and interpretable parameters for the spatial scale and smoothness
governed respectively by the parameters $r_{i}$ and $\nu_i$, the temporal scale $\alpha$, 
smoothness $a$ and the separability $b$. Note that the above class of models 
is fully symmetric, i.e. $C_{ij}(\textbf{h},u)=C_{ji}(\textbf{h},u) = C_{ij}(\textbf{h},-u) =C_{ij}(-\textbf{h},u)$ for all $i,j = 1,\dots,p$.

\section{Estimation}
\label{sec:estimation}

For a Gaussian random field, maximum full likelihood (FL) requires the evaluation of the determinant and the inverse of the $np\times np$ 
covariance matrix, where $n$ is the number of space-time points and $p$ the number of variables. 
The computational cost being of the order of $\mathcal{O}\left((np)^3\right)$ for both operations, a maximum full likelihood is 
unfeasible for large datasets, and even for datasets of moderate size as soon as the number of variables increases.

Composite likelihood (CL) methods \citep{Lindsay1988} have been proposed to perform efficient estimation with less time-consuming computational steps, and to 
guarantee good asymptotic properties. Composite likelihoods are products of smaller likelihoods defined on certain 
subsets of data such as marginal or conditional events, which are easy to compute.  
See \citet{Varin2011} for an overview. Following \citet{Bevilacqua2014}, we choose the pairwise marginal
 Gaussian likelihoods computed on all pairs of data $\{Z_i(\textbf{s}_\alpha,t_\alpha),
Z_j(\textbf{s}_\beta,t_\beta)\}$, where $i,j = 1,\dots,p$ and $\alpha, \beta = 1,\dots,n$, $n$ being the number of sites.
The negative log-likelihood of such pairs is 
\begin{equation*}
l(i,j,\textbf{s}_\alpha,\textbf{s}_\beta,t_\alpha,t_\beta;\theta)=
\frac{1}{2}\left\lbrace 
\log\Delta_{ij,\alpha\beta} + 
\frac{A_{ij,\alpha\beta}}
{\Delta_{ij,\alpha\beta}}
\right\rbrace
\end{equation*}
where $\Delta_{ij,\alpha\beta}=\sigma_i^2 \sigma_j^2-C_{ij}(\textbf{h},u)^2$ and 
$A_{ij,\alpha\beta}=\sigma_j^2Z_i(\textbf{s}_\alpha,t_\alpha)^2-2C_{ij}(\textbf{h},u)
Z_i(\textbf{s}_\alpha,t_\alpha)
Z_j(\textbf{s}_\beta,t_\beta)
+ \sigma_i^2Z_j(\textbf{s}_\beta,t_\beta)^2$, with 
$\textbf{h}=\|\textbf{s}_\alpha-\textbf{s}_\beta\|,u=|t_\alpha-t_\beta|$.
In the special case of multivariate symmetric models, the Weighted Pairwise log-Likelihood (WPL) is thus
\begin{align}
{\rm wpl}(\theta)=\sum_{(i,j,\alpha,\beta)\in \Lambda}
l(i,j,\textbf{s}_\alpha,\textbf{s}_\beta,t_\alpha,t_\beta;\theta)w_{\alpha\beta},\ \ \theta \in \Theta,
\label{eq:weighted composite likelihood}
\end{align}
where 
\begin{eqnarray*}
  \Lambda & = & \{ i=j=1,\dots,p,\quad \alpha=1,\dots,n-1,\quad  \beta=\alpha+1,\dots,n\} \\
          & \cup  & \{ i=1,\dots,p-1,\quad j=i+1,\dots,p,\quad \alpha,\beta=1,\dots,n\},
\end{eqnarray*}
and  $\Theta \subset \mathbb{R}^q$ is the space of parameters for which the model in (\ref{eq:new_model_final}) is valid, 
with $q=(p+2)(p+3)/2$ for the Gneiting-Mat\'ern class and $q=(p+2)(p+3)/2 +1$ for the Gneiting-Cauchy class.

\medskip

The computational cost of WPL is of the order of $\mathcal{O}\left((np)^2\right)$ when considering all possible pairs. It can be 
significantly reduced with an adequate choice of the weights. 
In this paper, we have chosen cut-off weights, that is $w_{\alpha\beta}=1$ if the space-time lag $(\| \textbf{h} \|,u) \leq (d_S,d_T)$, 
where the order is defined pointwise. Otherwise we set $w_{\alpha\beta}=0$. 
We use the expression \textit{weighted pairwise likelihood} but it is sometimes called \textit{truncated composite likelihood} 
or \textit{tapered composite likelihood}.
In spite of its simplicity, this weighting scheme provides a significant gain in computational time, and preserves a reasonable level of
statistical efficiency, 
since pairs of observations whose distance is beyond the correlation range are uninformative for the  smoothness and range parameters of the 
covariance function. The choice of $\textbf{d}=(d_S,d_T)$ will be discussed in details in the next section. At this stage,  we 
can notice that \textbf{d} is the same for all variables. 

\medskip

In the univariate setting, \citet{Bevilacqua2012} proposed to seek the ``optimal''  window $\textbf{d}^*$ such that
\begin{equation}
\textbf{d}^*=\arg \min_{\textbf{d}} {\rm tr}\left(\textbf{G}^{-1}_{np}(\textbf{d};\theta)\right),  \quad
\theta \in \Theta,
\label{eq:argmin}
\end{equation} 
where $\theta$ is replaced by a consistent estimator of $\theta$, such as for example the 
weighted least squares estimate based on empirical estimation of the variogram 
\citep[p. 91]{Cressie1993}. 
The value $\textbf{d}^*$ minimizes the sum of the diagonal elements of the inverse of
$\textbf{G}_{np}^{-1}(\textbf{d};\theta)$, which is 
the asymptotic variance-covariance matrix of the WPL estimators. 
The matrix $\textbf{G}_{np}(\textbf{d};\theta)$ is the Godambe information matrix, also referred to as the ``sandwich'' information matrix
\citep{Bevilacqua2012}.
The Godambe information matrix of a full log-likelihood function is equal to the Fisher information matrix.

\medskip

Estimating the parameters from model (\ref{eq:new_model_final}) requires to maximize
${\rm wpl}(\theta)$ in $\Theta$, when ${\rm wpl}(\theta)$ is computed with the appropriate window $\textbf{d}^*$.
Closed form expressions for the theoretical 
expression of the Godambe matrix are not easily feasible since first and second order derivatives 
of (\ref{eq:weighted composite likelihood}) with respect to all parameters are required.  We thus favor 
an empirical approach, based on a simulation study detailed in the next Section. 
For a given value of $\textbf{d}^*$, maximizing ${\rm wpl}(\theta)$ in $\Theta$ is by no means trivial, since  model in (\ref{eq:new_model_final})
has $(p+2)(p+3)/2$ parameters. Employing blindly an optimization function is deemed to fail, even for the simplest case with $p=2$.

\medskip

As a way to alleviate this problem,  ${\rm wpl}(\theta)$ is maximized sequentially in subspaces of $\Theta$,  
corresponding to blocks of related parameters, while keeping all other parameters fixed 
to the previously attained values.
Among the schemes that have been tested, the following combination seemed to lead to the best estimates:
for a fixed value of $b$, first seek the maximum corresponding to the $p(p-1)/2$ correlation parameters $\beta_{ij}$; 
then, for each variable $i$, find the maximum for $\sigma_i$, followed by $(\nu_i,r_i)$; 
finally, maximize with respect to the two temporal parameters $a$ and $\alpha$.
One iteration is achieved when all parameters have been estimated once. The likelihood is iteratively maximized 
until a stopping criterion is reached. We have chosen to stop the maximization whenever the log-likelihood is increased by less than one unit.
Other orders have been tested, leading to very close estimates. 

The separability parameter $b$, crucial in this study, is difficult to optimize in the above framework. We often
observed unstable maximization and/or estimates reaching the boundary of the parameter space $\Theta$. Since $b$ belongs to the interval
$[0,1]$, we decided to estimate it by a simple grid search. The parameter  $b$ is successively fixed to the values $k/10$, 
with $k=0,\dots,10$. For each of these values, the above maximization procedure is performed. Finally $\hat b$ is that value of $b$ 
corresponding to the highest maximized likelihood.
 
To perform the maximization within the subspaces of $\Theta$, 
we used the function \textit{optim} with quasi Newton method "BFGS" (published simultaneously in 1970 by Broyden, 
Fletcher, Goldfarb and Shanno) as implemented in R. Relevant initial values were set by estimating parameters of the corresponding 
marginal  spatial and temporal covariance functions, independently for each variable. Initial values of the correlation coefficients were 
set to their empirical values.

\section{Simulation study}
\label{sec:simulation}

We propose a simulation study with three main goals: setting the optimal window $\textbf{d}^*$, assessing the 
performance of the estimation procedure detailed in Section \ref{sec:estimation}, and measuring the gain in prediction
implied by model (\ref{eq:new_model_final}), as compared to 
non-separable or less flexible models. Because of the wide popularity of the Mat\'ern covariance functions, and also for space considerations,
we only report results obtained with the multivariate Gneiting-Mat\'ern class (\ref{eq:new_model_Matern}) in Theorem \ref{theo:GM}. 
The setting of the simulation study mimics the conditions found for the analysis of weather data as detailed in Section \ref{sec:data}. 
Three variables are available during one month (30 days) at 13 locations  (the precise locations are shown later, 
in Figure \ref{fig:map}). Among these,
11 sites are used for the first two tasks (circles)  while two sites are kept for validation (stars). 
There are 10 years of data, assumed independent. 
We thus simulate a sample of size 10 from a Gaussian vector with $30\times 13\times 3=1170$ 
components. The elements of the $1170 \times 1170$ matrix are given by the multivariate
space-time Mat\'ern covariance function with unit variance for all variables, i.e.
\begin{equation}
C_{ij}(\textbf{h},u)=\frac{\rho_{ij}}{\alpha |u|^{2a} +1}\ {\cal M}\left(\frac{\textbf{h}}{(\alpha |u|^{2a} +1)^{b/2}}; r_{ij},\nu_{ij}\right) \quad
i,j=1,\dots,p. 
\label{eq:model_simulation}
\end{equation}
Exact simulations with Cholesky decomposition are feasible, and a total of 100 repetitions are simulated. 
The reasonable dimension of the matrices ($1170\times 1170$) allows to compare WPL inference with a usual 
likelihood approach, referred to as the Full Likelihood (FL) approach.
Parameters have been chosen to match those estimated by an exploratory analysis of the spatial and temporal 
margins of the  dataset (see Table \ref{tab:estimation}). In order to account for
lack of space-time separability, the parameter $b$ was set equal to 0.8. As already pointed out earlier, we only report results obtained 
with the Mat\'ern model.

\subsection{Setting the optimal window}

The optimal window $\textbf{d}^*$ given in Eq. (\ref{eq:argmin}) minimizes the sum of the estimation variance of every parameter. 
For several choices of the spatial distance $d_S \in\{250, 500, 750\}$ (expressed in kms), and temporal distance $d_T \in \{2,5,10\}$ (in days),  
the estimation of the parameters is performed on each 
simulated repetition. Then, empirical estimates of the estimation variances are computed. Their sum is an estimate of the criterion in
Eq. (\ref{eq:argmin}), as recommended in \citet{Bevilacqua2012}. Results are reported in Table \ref{tab:distance_choice}.  
As the spatial distance increases, the quantity $\textbf{G}_{np}^{-1}(\textbf{d};\hat{\theta})$
 increases for all temporal distances. 
For all spatial distances, the criterion is minimal for the smallest temporal window. 
Parameters are best estimated if pairs corresponding to moderate distances and to successive days only are used,
which corresponds to $7.2\%$ and $13.5\%$ of the number of pairs for the two lowest values of the criterion. 
 Overall, the window
$\textbf{d} =(250{\rm km},2{\rm days})$ is optimal according to this criterion. The criterion computed with 
$\textbf{d} =(500{\rm km},2{\rm days})$ is very close to the optimal.
The estimated mean and variance of the separability parameter $b$ for the different space-time windows,  is 
reported in Table \ref{tab:separability_parameter}. The picture is now slightly different. Bias and estimation variance are minimal for 
$\textbf{d} =(500{\rm km},2{\rm days})$. Considering the importance of the separability parameter, and since the trace criterion 
for this window is very close to the optimum, the window is set to $\textbf{d} =(500{\rm km},2{\rm days})$ for the rest of this work.
These results are 
consistent with those reported in \citet{Bevilacqua2012}, where it is observed that the relative efficiency of 
WPL as a function of the distance first increases to a maximum and then decreases as more distant pairs are added in the WPL.

\begin{table}
\centering
\begin{tabular}{c|cccccc}
\hline \hline
{\bf d} & \multicolumn{2}{c}{$d_T$ = 2 days} & \multicolumn{2}{c}{$d_T$ = 5 days} & \multicolumn{2}{c}{$d_T$ = 10 days} \\
\hline
 $d_S$ = 250 km & {\bf 0.120} & (7.2\%) & 0.148 & (15.1\%) & 0.185 & (26.2\%) \\
 $d_S$ = 500 km & 0.127 & (13.5\%) & 0.162 & (28.3\%) & 0.203 & (49.1\%) \\
 $d_S$ = 750 km & 0.146 & (15.4\%) & 0.196 & (33.3\%) & 0.227 & (57.7\%) \\
 \hline \hline
\end{tabular}
\caption{Average of the sum of the estimated variances of all parameters for several  windows $\textbf{d}=(d_S,d_T)$.
100 synthetic datasets simulated according to model (\ref{eq:model_simulation}).
The number in brackets corresponds to the percentage of pairs used in the computation of {\rm wpl}.\label{tab:distance_choice}}
\end{table}

\begin{table}
\centering
\begin{tabular}{c|cc|cc|cc}
\hline\hline
{\bf d} & \multicolumn{2}{c|}{$d_T$ = 2 days} & \multicolumn{2}{c|}{$d_T$ = 5 days} & \multicolumn{2}{c}{$d_T$ = 10 days} \\
\hline
 $d_S$ = 250 km & 0.694 & (0.234) & 0.681 & (0.268) & 0.630 & (0.305) \\
 $d_S$ = 500 km & {\bf 0.768} & ({\bf 0.230}) & 0.731 & (0.251) & 0.717 & (0.259) \\
 $d_S$ = 750 km & 0.744 & (0.245) & 0.723 & (0.257) & 0.696 & (0.281) \\
 \hline\hline
\end{tabular}
\caption{Estimated mean and standard deviation (in brackets) of the separability parameter $b$ for several  
windows $\textbf{d}=(d_S,d_T)$. Same simulations as in Table \ref{tab:distance_choice}.\label{tab:separability_parameter}}
\end{table}

\subsection{Efficiency of the estimation procedure}

We now assess the ability of the estimation procedure to estimate efficiently the different parameters, the window being set
to $\textbf{d} =(500{\rm km},2{\rm days})$. This simulation study also allows us to compare the efficiency of WPL
with respect to FL for the multivariate Gneiting-Mat\'ern class. We maximize WPL and FL, using the same maximization scheme, on the same set of 100 simulations. 
As can be observed from Table \ref{tab:estimation}, the overall performances are good for both approaches.
For most parameters, the difference between the median and the mean is negligible, and differences between the mean and the true value
is relatively small for most parameters. The range and smoothness parameters are a notable exception. 
Their estimators are less biased and more dispersed when using WPL than when using FL.  
For all variables $i\in \{1,2,3\}$,  the smoothness parameters, $\nu_i$, tend to be overestimated, while the 
scale parameters, $1/r_i$, tend to be underestimated.  In \citet{Zhang2004}, it is shown  that these parameters compensate 
each other and that the simultaneous estimation of both is difficult.
Results shown in Table \ref{tab:estimation} confirm our assertions. In accordance with the parameterization chosen in Theorems 
\ref{theo:GM} and \ref{theo:GC}, the correlation coefficients $\beta_{ij}$, with $1 \leq i < j \leq 3$ must belong to the 
interval $(-1,1)$ with $\beta_{ii}=1$, for $i=1,\dots,3$, and the matrix 
$[\beta_{ij}]_{i,j=1}^3$ must be positive definite.
The estimators of the correlations coefficients $\beta_{ij}$, with $1 \leq i < j \leq 3$, are unbiased for both approaches,
 but their dispersion is much larger when using WPL than when using  FL. We return to this point later.
Remember that the separability parameter $b$ is not estimated continuously and 
that the window \textbf{d} has been chosen to minimize the bias and the variance of its estimator. Interestingly, it is better estimated using WPL than 
using FL.  With WPL, the median is equal to the true value and it is close to the mean. The separability 
parameter is estimated to be non null, for all simulated datasets but one.

\medskip

To complete this comparison, we compute the relative efficiency of WPL with respect to FL. For each approach, and
for each of the 15 parameters $\theta_i$, with $i \leq i \leq 15$, the root mean square error, 
${\rm rmse}_{i} = \sqrt{{\rm bs}_i^ 2 + {\rm sd}_i^ 2}$, is computed, where the bias is 
${\rm bs}_i = \bar{\hat{\theta_i}} - \theta_i$, and the variance is 
${\rm sd}_i^2=  \sum_{j=1}^{100}(\hat{\theta}_{j,i}-\bar{\hat{\theta}}_i)^2/100$, with
$\bar{\hat{\theta_i}}=\sum_{j=1}^{100} \hat{\theta}_{j,i}/100$. For each variable, we denote ${\rm rmse}_{i}^{\rm FL}$ and ${\rm rmse}_{i}^{\rm WPL}$
the root mean square error corresponding respectively to FL  and to WPL. 
The root relative efficiency, defined as 
$${\rm rre}_i = {\rm rmse}^{\rm FL}_i / {\rm rmse}^{\rm WPL}_i,$$ 
is reported for each variable in Table \ref{tab:root_relative_effiency}.  When $\hbox{rre} < 1$,
FL is more efficient than WPL, and conversely when $\hbox{rre}>1$. 
For all parameters, it was found that FL leads to higher bias than WPL, but that the lower 
variance associated to FL usually more than compensates this effect. On these simulations, WPL is found to be rather inefficient 
for the estimation of the correlation parameters $\beta_{ij}$, for which the variance associated to WPL is about 30 times that 
associated to FL. For all other parameters and in particular for regularity and range parameters, WPL provides estimates with a mild 
loss in efficiency as compared to a full likelihood approach, with a significant gain in terms of computation. 
Similar results were obtained for the inference of the parameters of Max-Stable processes in \citet{Castruccio2014}. Rarely, 
the relative efficiency is even slightly larger than 1 but this could be due to the moderate number of replicates. 
In these cases, the lower variance for FL does not compensate the larger bias.

\begin{table}
\centering
{\scriptsize
\begin{tabular}{r|r|rrrrrr|rrrrrr}
\hline\hline
 &  &  \multicolumn{6}{c|}{Weighted Pairwise Likelihood} & \multicolumn{6}{c}{Full Likelihood} \\
 & True & Min & $Q_1$ & Median & Mean & $Q_3$ & Max & Min & $Q_1$ & Median & Mean & $Q_3$ & Max \\
\hline
\hline
$\sigma_1$ & 1.00 & 0.89 & 0.97 & 1.00 & 1.00 & 1.02 & 1.16 & 0.93 & 0.97 & 0.99 & 0.99 & 1.01 & 1.06 \\
$\sigma_2$ & 1.00 & 0.91 & 0.97 & 1.00 & 1.00 & 1.02 & 1.11 & 0.95 & 0.98 & 0.99 & 1.00 & 1.01 & 1.06 \\
$\sigma_3$ & 1.00 & 0.92 & 0.97 & 0.99 & 0.99 & 1.02 & 1.10 & 0.94 & 0.97 & 0.99 & 0.99 & 1.00 & 1.12 \\
\hline
$\beta_{12}$ & $-0.40$ & $-0.61$ & $-0.46$ & $-0.40$ & $-0.4$0 & $-0.34$ & $-0.1$5 & $-0.45$ & $-0.42$ & $-0.41$ & $-0.41$ & $-0.40$ & $-0.37$ \\
$\beta_{13}$ & $-0.40$ &$ -0.62$ & $-0.47$ & $-0.41$ & $-0.41$ & $-0.34$ & $-0.21$ & $-0.45$ & $-0.42$ & $-0.41$ & $-0.41$ & $-0.40$ & $-0.36$ \\
$\beta_{23}$ & 0.25 & $-0.05$ & 0.19 & 0.27 & 0.26 & 0.33 & 0.47 & 0.21 & 0.25 & 0.26 & 0.26 & 0.27 & 0.31 \\
\hline
$\nu_1$ & 0.70 & 0.55 & 0.69 & 0.74 & 0.75 & 0.82 & 1.01 & 0.58 & 0.72 & 0.75 & 0.76 & 0.79 & 0.99 \\
$\nu_2$ & 0.80 & 0.55 & 0.76 & 0.83 & 0.84 & 0.91 & 1.17 & 0.68 & 0.82 & 0.87 & 0.88 & 0.95 & 1.17 \\
$\nu_3$ & 0.40 & 0.28 & 0.39 & 0.43 & 0.42 & 0.46 & 0.72 & 0.30 & 0.43 & 0.45 & 0.45 & 0.47 & 0.59 \\
\hline
$1/r_1$ & 250 & 135 & 205 & 229 & 235 & 262 & 388 & 147 & 199 & 216 & 220 & 238 & 333 \\
$1/r_2$ & 200 & 128 & 170 & 190 & 192 & 215 & 299 & 131 & 158 & 177 & 178 & 193 & 261 \\
$1/r_3$ & 350 & 149 & 284 & 310 & 327 & 356 & 1017 & 195 & 247 & 277 & 283 & 293 & 1017 \\
\hline
$\alpha$ & 0.90 & 0.65 & 0.84 & 0.90 & 0.91 & 0.99 & 1.24 & 0.78 & 0.91 & 0.95 & 0.95 & 1.00 & 1.12 \\
$a$ & 0.50 & 0.35 & 0.46 & 0.49 & 0.50 & 0.54 & 0.71 & 0.46 & 0.49 & 0.51 & 0.51 & 0.53 & 0.58 \\
$b$ & 0.80 & 0.00 & 0.60 & 0.80 & 0.77 & 1.00 & 1.00 & 0.30 & 0.50 & 0.60 & 0.64 & 0.70 & 1.00 \\
\hline\hline

\end{tabular}
}
\caption{Summary statistics of the estimated parameters for $\textbf{d} =(500\ {\rm km},2\ {\rm days})$.
Same simulations as in Table \ref{tab:distance_choice}.}
\label{tab:estimation}
\end{table}

\begin{table}
\centering
{\small
\begin{tabular}{c|cc|cc|cc|cc|cc|cc|}
\hline\hline
 & \multicolumn{2}{c|}{$\sigma_1$} & \multicolumn{2}{c|}{$\sigma_2$} & \multicolumn{2}{c|}{$\sigma_3$} 
 & \multicolumn{2}{c|}{$\beta_{12}$} & \multicolumn{2}{c|}{$\beta_{13}$} & \multicolumn{2}{c|}{$\beta_{23}$} \\
 & WPL & FL & WPL & FL & WPL & FL  & WPL & FL & WPL & FL & WPL & FL \\
\hline
${\rm bs}^2\ (\times 10^{-5})$ & 0.18 & 5.13 & 1.13 & 2.54 & 3.48 & 17.1 & 1.02 & 5.77 & 6.30 & 4.50 & 7.96 & 5.35  \\
\hline
${\rm sd}^2\ (\times 10^{-3})$ & 1.73 & 0.84 & 1.40 & 0.59 & 1.24 & 0.63 & 8.47 & 0.26 & 7.86 & 0.29 & 10.8 & 0.36 \\
\hline
rmse                           & 0.042& 0.030 & 0.038 & 0.025 & 0.036 & 0.028 &  0.092 & 0.018 & 0.089 & 0.018 & 0.104 & 0.020 \\
\hline
rre    & \multicolumn{2}{c|}{0.72} & \multicolumn{2}{c|}{0.66} & \multicolumn{2}{c|}{0.79} & \multicolumn{2}{c|}{0.19} & \multicolumn{2}{c|}{0.21} & \multicolumn{2}{c}{0.19} \\   
\hline
\hline
 & \multicolumn{2}{c|}{$\nu_1$} & \multicolumn{2}{c|}{$\nu_2$} & \multicolumn{2}{c|}{$\nu_3$} 
& \multicolumn{2}{c|}{$\alpha$} & \multicolumn{2}{c|}{a} & \multicolumn{2}{c|}{b} \\  
 & WPL & FL & WPL & FL & WPL & FL & WPL & FL & WPL & FL & WPL & FL \\
\hline
${\rm bs}^2\ (\times 10^{-3})$ & 2.63 & 3.48 & 1.75 & 6.30 & 0.60 & 2.95 & 0.132 & 2.82 & 0.001 & 0.018 & 1.000 & 27.0 \\
\hline
${\rm sd}^2\ (\times 10^{-3})$ & 9.71 & 4.30 & 13.1 & 8.95 & 3.81 & 1.82 & 11.4 & 4.70 & 5.08 & 0.699 & 52.7 & 19.4 \\
\hline
rmse                          & 0.111 & 0.088 & 0.122 & 0.123 & 0.066 & 0.069 & 0.107 & 0.087 & 0.071 & 0.027 &  0.232 & 0.215 \\
\hline
rre & \multicolumn{2}{c|}{0.79} & \multicolumn{2}{c|}{1.01} & \multicolumn{2}{c|}{1.04}
& \multicolumn{2}{c|}{0.81} & \multicolumn{2}{c|}{0.38} & \multicolumn{2}{c|}{0.93}\\     
\hline\hline
 & \multicolumn{2}{c|}{$1/r_1$} & \multicolumn{2}{c|}{$1/r_2$} & \multicolumn{2}{c|}{$1/r_3$} & & & & & & \\ 
 & WPL & FL & WPL & FL & WPL & FL & & & & & & \\
\hline
${\rm bs}^2$ & 218.5 & 911.6 & 61.8 & 473.9 & 516.6 & 4539.3 & & & & & & \\
\hline
${\rm sd}^2$ & 2300.0 & 856.1 & 1127.1 & 682.4 & 9629 & 6788.9 & & & & & & \\
\hline
rmse         & 50.2   & 42.0  & 34.5 & 34.0 & 100.7 & 106.4 & & & & & & \\
\hline
rre & \multicolumn{2}{c|}{0.84} & \multicolumn{2}{c|}{0.99} & \multicolumn{2}{c|}{1.06} & & & & & & \\   
\hline
\hline
\end{tabular}}
\caption{Bias (bs), variance (sd$^2$), root mean square error (rmse) and root relative efficiency (rre) for each parameter. 
Same simulations as in Table \ref{tab:distance_choice}. \label{tab:root_relative_effiency}}
\end{table}

\subsection{Assessing the predictive performances}

Finally, the predictive performance of the model is assessed. In addition to the non-separable 
model according to which data are simulated, we also consider 
three more specific models, which are special cases of the full model. We compare space-time separable (S) and non-separable (NS) models, 
and models with equal (E) or different (D) smoothness and scale parameters. We will thus consider the following 4 models:
\begin{enumerate}
\item S-E: $b=0$; all components have equal smoothness and scale parameters. The covariance matrix follows
(\ref{eq:full_sep}).

\item NS-E: $0<b\leq 1$; all components have equal smoothness and scale parameters. This model is space-time non-separable, but it has the 
same space-time covariance for all variables. The covariance matrix-valued function follows (\ref{eq:space_time_sep}).

\item S-D: $b=0$; each component has its own smoothness and scale parameters, but it is space-time separable.

\item NS-D: $0<b\leq 1$; each component has its own smoothness and scale parameters, and it is not  space-time separable. It is  
the non-separable model defined through Eq. (\ref{eq:model_simulation}).
\end{enumerate}

Data are simulated according to model NS-D, with parameters as in Table \ref{tab:estimation},
at the 13 stations displayed in Figure \ref{fig:map}. Two stations, indicated with a star, have been selected 
for validation. Data at these two stations are not used to estimate the parameters. For each day $t$, the conditional distribution 
is computed at these two locations, given the sets $\{$all data at time $t-1,\ldots,t-q\} \cup \{$data at other locations 
at the same time $t \}$, where $q \in \{2,3,\dots\}$ is the number of days used for the prediction. 
Under Gaussianity, this amounts to compute the $6-$variate conditional expectation and the conditional covariance matrix.

The four models are compared by means of four different scores: Root Mean Square Error (RMSE), Mean Absolute Error (MAE), 
the Continuous Ranked Probability Score (CRPS) and Logarithmic Score (LogS) 
\citep{GneitingRaftery2007}. Let use denote $\tilde{z}^t_1,\dots,\tilde{z}^t_6$ the conditional expectation
of the 6 predicted variables (2 sites $\times$ 3 variables) for a given day $t$, $\tilde{\sigma}_1,\dots,\tilde{\sigma}_6$ the corresponding conditional standard deviations,
$z^t_1,\dots,z^t_6$ being the observed values.  Note that the conditional standard deviations are independent of the day $t$, since they only depend
on covariance values.

The first two scores, MAE and RMSE, compare the conditional expectation to the true value. The MAE is defined as
$${\rm MAE} = \frac{1}{6|T|} \sum_{t \in T} \sum_{j=1}^6 |\tilde{z}^t_j - z^t_j|,$$
where the sum is taken over the set of all testing days, $T$ and $|T|$ denotes the cardinality of the set $T$. 
The mean square error (MSE) is
$${\rm MSE} = \frac{1}{6|T|} \sum_{t \in T} \sum_{j=1}^6 (\tilde{z}^t_j - z^t_j)^2.$$
The RMSE is the square root of the MSE and has the advantage of being recorded in the same unit as the data.

The  other two scores, CRPS and LogS, asses not only the prediction but its variance as well. They are easily computed in
the case of a normal predictive distribution. 
The CRPS measures the discrepancy between the predictive cumulative distribution function (CDF) and the true value. Specifically, if $F$ is the predictive 
CDF and $z$ the true value, crps is defined as
$${\rm crps} (F,z) = \int_{-\infty}^\infty [F(u) - H(u-z)]^2 du,$$
where $H(u-z)$ denotes the Heaviside function, which takes the value 0 when $u < z$ and 1 otherwise. 
In case of normal CDF $\Phi_j^t$ with expectation
$\tilde{z}^t_j$ and variance $\tilde{\sigma}^2_j$, one can show that
$${\rm crps} (\Phi_j^t,z) = \tilde{\sigma}^t_j\{ \tilde z [2 \Phi( \tilde z) -1  ] + 2 \phi(\tilde z) - \pi^{-1/2}\},$$
where $\tilde{z} = (z - \tilde{z}^t_j)/ \tilde{\sigma}_j$ is the normalized prediction error and where
$\phi(\cdot)$ and $\Phi(\cdot)$ denote the density and the CDF, respectively, of the normal distribution with mean 0 and variance 1. The CRPS is
$$ {\rm CRPS} = \frac{1}{6|T|} \sum_{t \in T} \sum_{j=1}^6 {\rm crps} (\Phi_j^t,y_j^t).$$
It is easy to show that the CRPS tends to the MAE when  $\tilde{\sigma_j} \to 0$, for all $j=1,\dots,6$.  For this reason, the CRPS can be interpreted as a generalized 
version of the MAE \citep{GneitingRaftery2007}.
The marginal logarithmic score is the negative of the logarithm of the marginal predictive density at the true value
$${\rm LogS}_1 = \frac{1}{6|T|} \sum_{t \in T} \sum_{j=1}^6  \left\{ \ln \tilde{\sigma}_j  + \frac{(\tilde{z}^t_j - z^t_j)^2}{2 \tilde{\sigma}^2_j} \right\}.$$
The multivariate logarithmic score considers the multivariate predictive density computed at the true vector $\textbf{z}^t = (z_1,\dots,z_6)^\top$
$${\rm LogS}_6 = \frac{1}{6|T|} \sum_{t \in T}   \left\{ \ln \det  \tilde{\mathbf{\Sigma}}^t +  
\frac{ (\tilde{\textbf{z}}^t - \textbf{z}^t)^\top \tilde{\mathbf{\Sigma}}^{-1} (\tilde{\textbf{z}}^t - \textbf{z}^t )}{2}  \right\},$$
where $\tilde{\mathbf{\Sigma}}$ is the conditional covariance matrix and $\textbf{a}^\top$ is the transpose of $\textbf{a}$.

\medskip

\begin{figure}
\includegraphics[width = 15cm, height = 12cm]{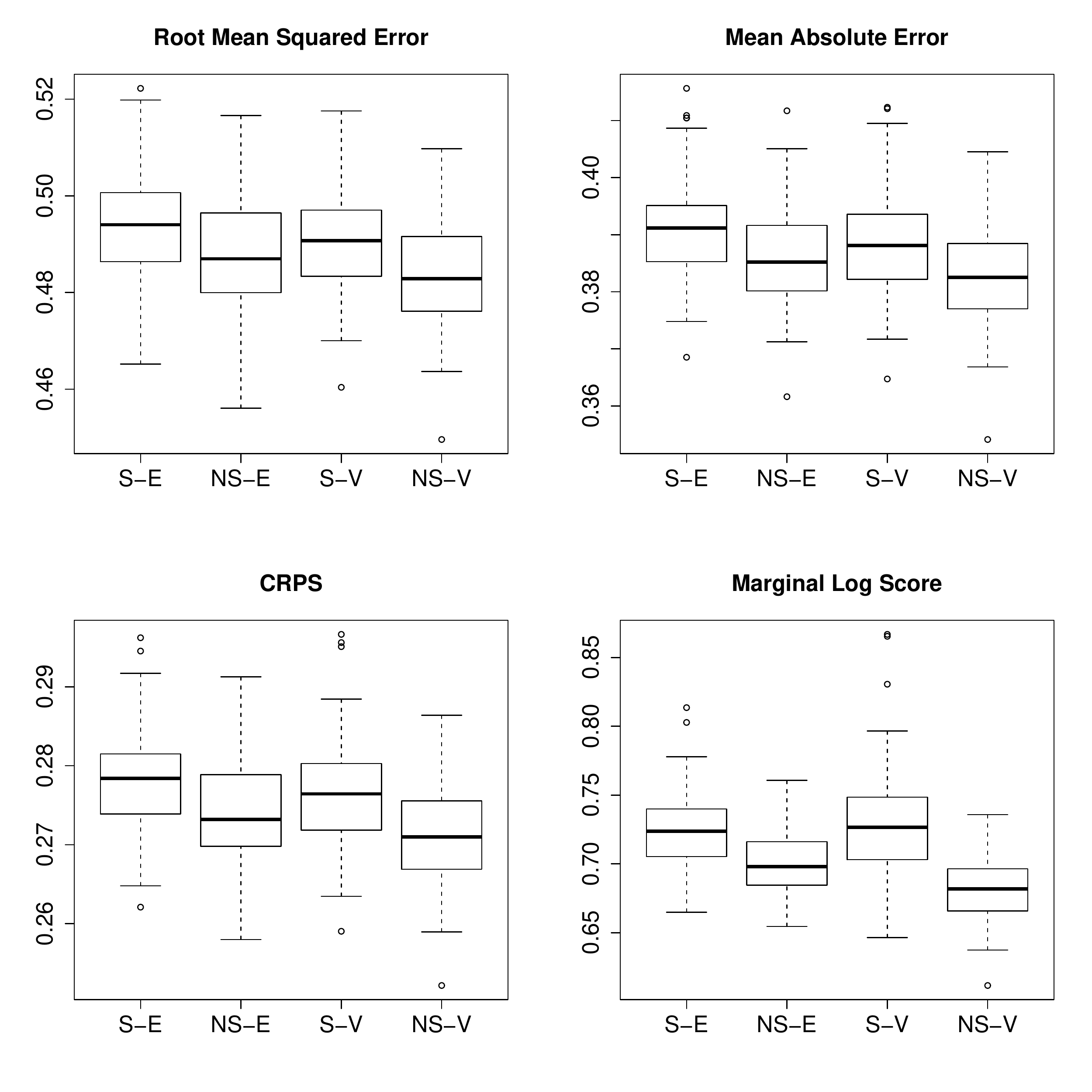}
\caption{Prediction scores RMSE, MAE, CRPS and LogS$_1$ for the  models S-E, NS-E, S-D and NS-D.  Same simulations as in Table \ref{tab:distance_choice}. Estimates 
obtained with WPL.
\label{fig:boxplot}}

\end{figure}

The scores defined above are computed for the set of 100 simulations, for each of the four models,
and for both estimation approaches, WPL and FL. The average scores are reported in Table \ref{tab:pred_score}, and 
the boxplots for WPL are shown in Figure \ref{fig:boxplot}. We only 
report results obtained using the two previous days as conditioning data (i.e. $q=2$),
but similar results were obtained for $q=1$ and $q=3$. 
Recall that lower scores indicate a better adequacy between the model and the data.

Some comments are in order: in general, 
differences in score values are low, but they are consistently observed on all simulations. Therefore, they can be considered as 
significant. Obviously, the highest scores correspond to the fully
separable model, SE,  whereas the lowest scores are obtained for the non-separable model, NS-D, for which the difference between WPL and FL is 
negligible. Differences between models are more  pronounced when considering scores that also involve the conditional variance (CRPS and LogS).
Accounting for non-separability  makes a more important difference than having different smoothness and scale parameters for each variable.
It is interesting to note that the difference between the models is more pronounced with WPL than with FL. 

In conclusion, this simulation study shows that WPL is a valid procedure for estimating the parameters of the Gneiting-Mat\'ern multivariate space-time model
when analyzing large  data sets for which a Full Likelihood maximization is not feasible. The fact that the difference in scores between the 
models is more pronounced with WPL than with FL is a strong indication  that it is particularly interesting to apply WPL to the  
non-separable NS-D model. When selecting the correct model, we note that in terms of prediction performances (RMSE, MAE, CRPS), the differences are negligible 
when using WPL estimates instead of FL estimates despite the lower estimating efficiency shown in Table \ref{tab:root_relative_effiency}. 

\begin{table}
\centering
\begin{tabular}{l|ccccc|ccccc}
\hline \hline
 & \multicolumn{5}{c|}{Weighted Pairwise Likelihood} & \multicolumn{5}{c}{Full Likelihood} \\
 & RMSE & MAE & CRPS & LogS$_1$ & LogS$_6$ & RMSE & MAE & CRPS & LogS$_1$ & LogS$_6$ \\
\hline
 S-E  & 0.494 & 0.391 & 0.278 & 0.725 & 4.070 & 0.489 & 0.386 & 0.274 & 0.690 & 3.792 \\
 NS-E & 0.488 & 0.386 & 0.274 & 0.701 & 3.903 & 0.488 & 0.386 & 0.273 & 0.688 & 3.782 \\
 S-D  & 0.490 & 0.389 & 0.277 & 0.729 & 4.091 & 0.484 & 0.383 & 0.271 & 0.680 & 3.733 \\
 NS-D & {\bf 0.485} & {\bf 0.383} & {\bf 0.271} & {\bf 0.683} & {\bf 3.789} &{\bf 0.484} & {\bf 0.383} & {\bf 0.271} & {\bf 0.678} & {\bf 3.723}\\
 \hline\hline
\end{tabular}
\caption{Mean of the prediction scores RMSE, MAE, CRPS, LogS$_1$ and  LogS$_6$,  according to models S-E, NS-E, S-D and NS-D, and for 
WPL and FL approaches. Same simulations as in Table \ref{tab:distance_choice}
\label{tab:pred_score}.}
\end{table}

\section{Western France weather dataset}
\label{sec:data}

\begin{figure}[ht]
  \begin{center}
\includegraphics[width = 15cm]{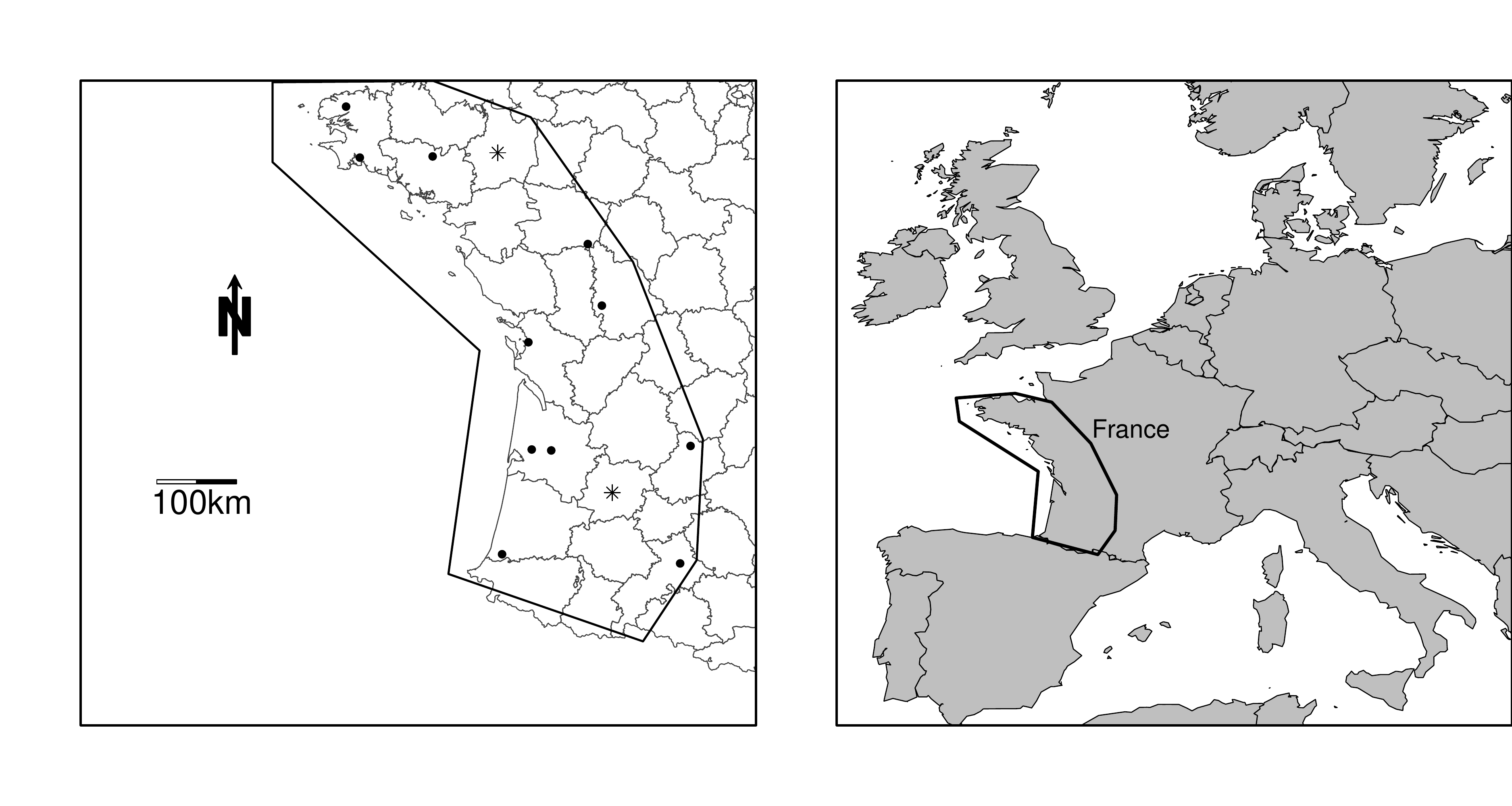}    
  \end{center}
\caption{Location of the 13 weather stations over western France. Stars: validation stations.\label{fig:map}}
\end{figure}

This section illustrates the use of the Gneiting-Mat\'ern multivariate space-time model for the analysis of a weather dataset which consists of three daily variables 
(solar radiation, R, temperature, T, and humidity, H) recorded at 13 stations in Western France from 2003 to 2012 (see Figure \ref{fig:map}). 
These data are part of the French National Institute for Agricultural Research (INRA) archive of weather data. 
The considered domain experiences an oceanic climate, characterized 
by moist and cool (but not cold) winters. Due to the prevailing westerly winds,  some amount of space-time interaction 
is thus expected, from West to East. The validation stations (indicated with stars in Figure \ref{fig:map})
have been selected in the Eastern part of the domain, neither too close,  nor too far from the other stations. The first validation station, Le Rheu, is located near Rennes,
in Brittany. The second one, Bourran, is located near Agen, in Aquitaine.  The  other 11 stations are used to estimate the parameters. In order 
to restrict ourselves to data being stationary in time, we selected data recorded in January, from 2003 to 2012.

\begin{figure}
\includegraphics[width = 15cm]{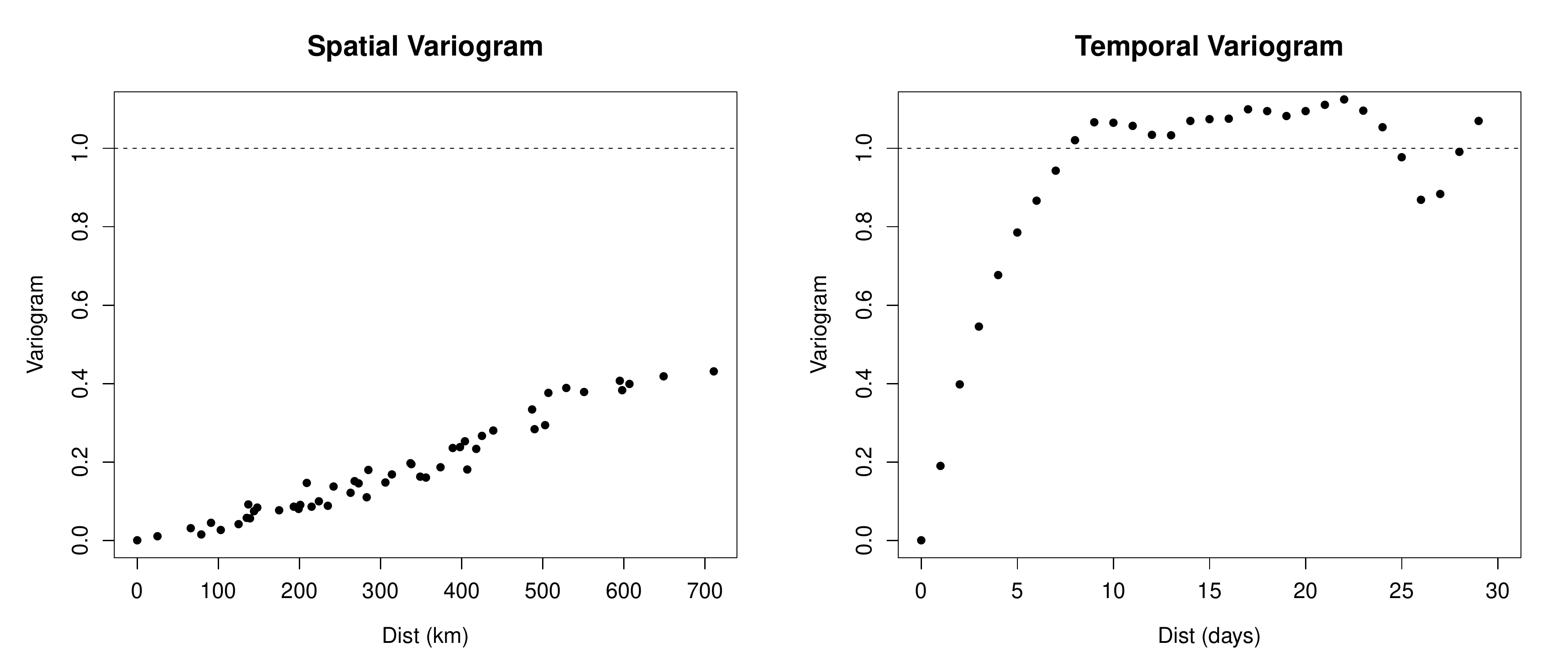}
\centering
\caption{Marginal spatial and temporal variograms for daily January temperature standardized variables.\label{fig:temp}}

\end{figure}

To model the January data 
$\textbf{Y}(\textbf{s},t)=\{Y_{\rm R}(\textbf{s},t),Y_{\rm T}(\textbf{s},t),Y_{\rm H}(\textbf{s},t)\}^\top$, we consider a model that is stationary in time and non-stationary in space, namely
\begin{equation}
Y_i(\textbf{s},t)=\mu_i(\textbf{s})+\sigma_i(\textbf{s})
Z_i(\textbf{s},t)\quad i \in\{{\rm R,T,H} \} \quad 
(\textbf{s},t)\in\mathbb{R}^2\times\mathbb{R}.
\end{equation}
The data are first standardized at each location by their averages and standard deviations. 
The space-time modeling will be carried out on the standardized variables, 
$$\textbf{Z}(\textbf{s},t)=\{Z_{\rm R}(\textbf{s},t),Z_{\rm T}(\textbf{s},t),Z_{\rm H}(\textbf{s},t)\}^\top,$$
which are assumed to be space-time stationary and multivariate Gaussian. The simultaneous observation of the spatial and temporal variograms
of the daily temperature standardized variables,
depicted in Figure \ref{fig:temp}, reveals a zonal anisotropy between space and time. While the temporal variogram reaches a sill equal to the overall 
variance, equal to 1 due to the standardization, the spatial variogram does not exceed 0.4, even at very large distances. This zonal anisotropy is 
modeled by considering that the process $\textbf{Z}$ arises as the sum of a zero mean multivariate temporal Gaussian process \textbf{X} accounting for 
temporal effects being constant in space, and an independent zero mean multivariate space-time Gaussian process \textbf{W}
\begin{equation}
Z_i(\textbf{s},t)=X_i(t)+W_i(\textbf{s},t)\quad i \in\{{\rm R,T,H} \}, \quad 
(\textbf{s},t)\in\mathbb{R}^2\times\mathbb{R}.
\label{eq:sumProcesses}
\end{equation}
Physically speaking, the model (\ref{eq:sumProcesses}) assumes that part of the daily changes in the weather pattern, which accounts for 60\% of the overall variability,
is regional and affects the whole domain under study. The stationary space-time variations account for 40\% of the variability.

This model provides the flexibility to accurately model the zonal anisotropy observed in Figure \ref{fig:temp}, since 
\begin{eqnarray*}
\gamma_{ii}^{(\textbf{Z})}(\textbf{h},0)&=&\gamma_{ii}^{(\textbf{X})}(0)+\gamma_{ii}^{(\textbf{W})}(\textbf{h},0)\underset{\textbf{h} \to \infty}{\longrightarrow}
\left[\sigma_i^{(\textbf{W})}\right]^2\\
\gamma_{ii}^{(\textbf{Z})}(\textbf{0},u)&=&\gamma_{ii}^{(\textbf{X})}(u)+\gamma_{ii}^{(\textbf{W})}(\textbf{0},u)\underset{u \to \infty}{\longrightarrow}
\left[\sigma_i^{(\textbf{X})}\right]^2 +\left[\sigma_i^{(\textbf{W})}\right]^2 =1.
\end{eqnarray*}
We further suppose that the process \textbf{W} is Gaussian, with a Gneiting-Mat\'ern covariance function, as described in Eq. 
(\ref{eq:new_model_Matern})
and that the temporal process \textbf{X} has covariance function 
$C_{ij}^{(\textbf{X})}(u)=
\sigma_i^{(\textbf{X})} \sigma_j^{(\textbf{X})}\beta_{ij}^{(\textbf{X})}/(\alpha^{(\textbf{X})}|u|^{2 a^{(\textbf{X})}}+1)$,
 with $i,j \in \{{\rm R,T,H} \}$ and $u \in \mathbb{R}$. Note that this temporal covariance function is the temporal marginal of a 
Gneiting-Mat\'ern covariance (i.e. when $\textbf{h}=\textbf{0}$.

\medskip

\begin{figure}
  \begin{center}
    \includegraphics[width=7.5cm]{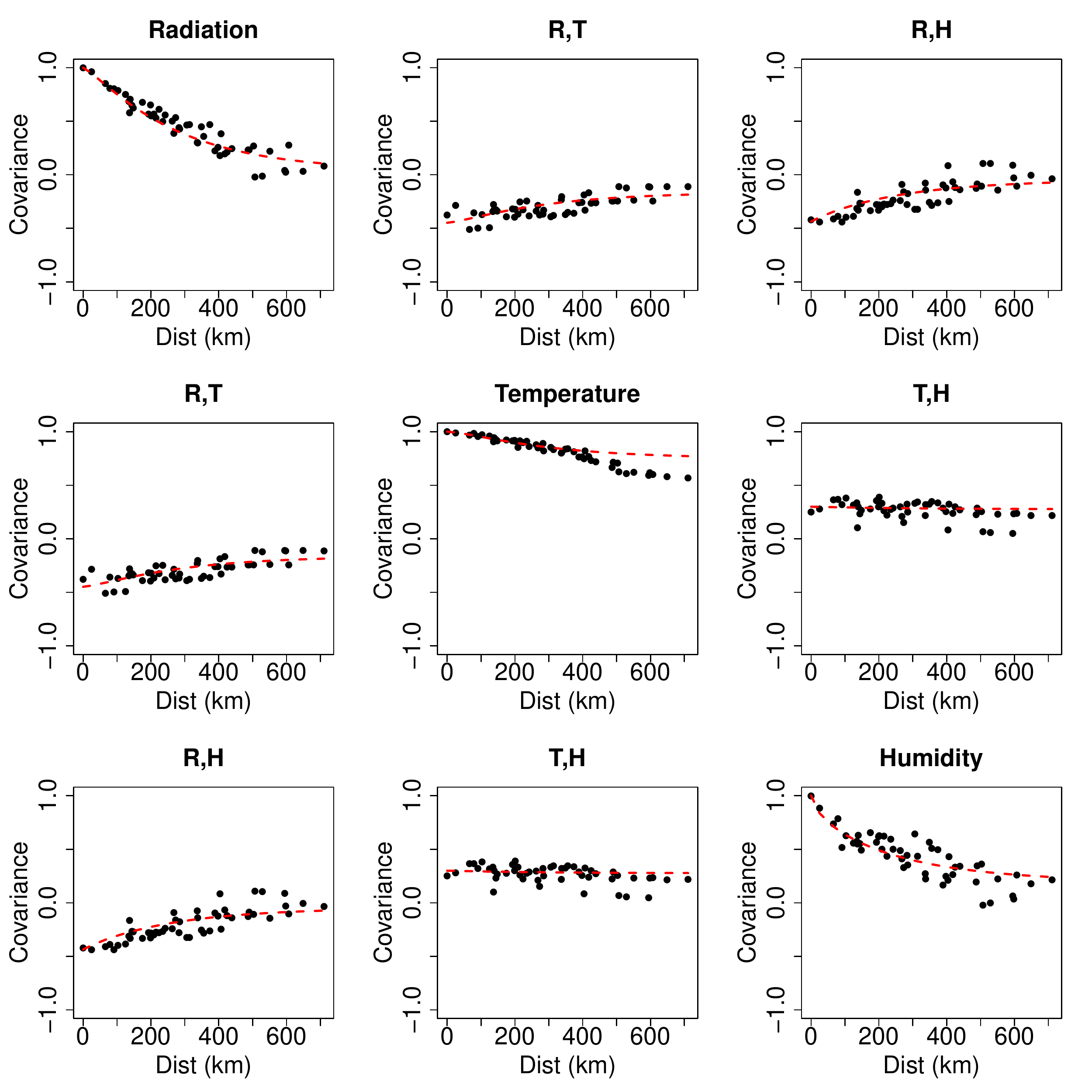}
\quad
 \includegraphics[width=7.5cm]{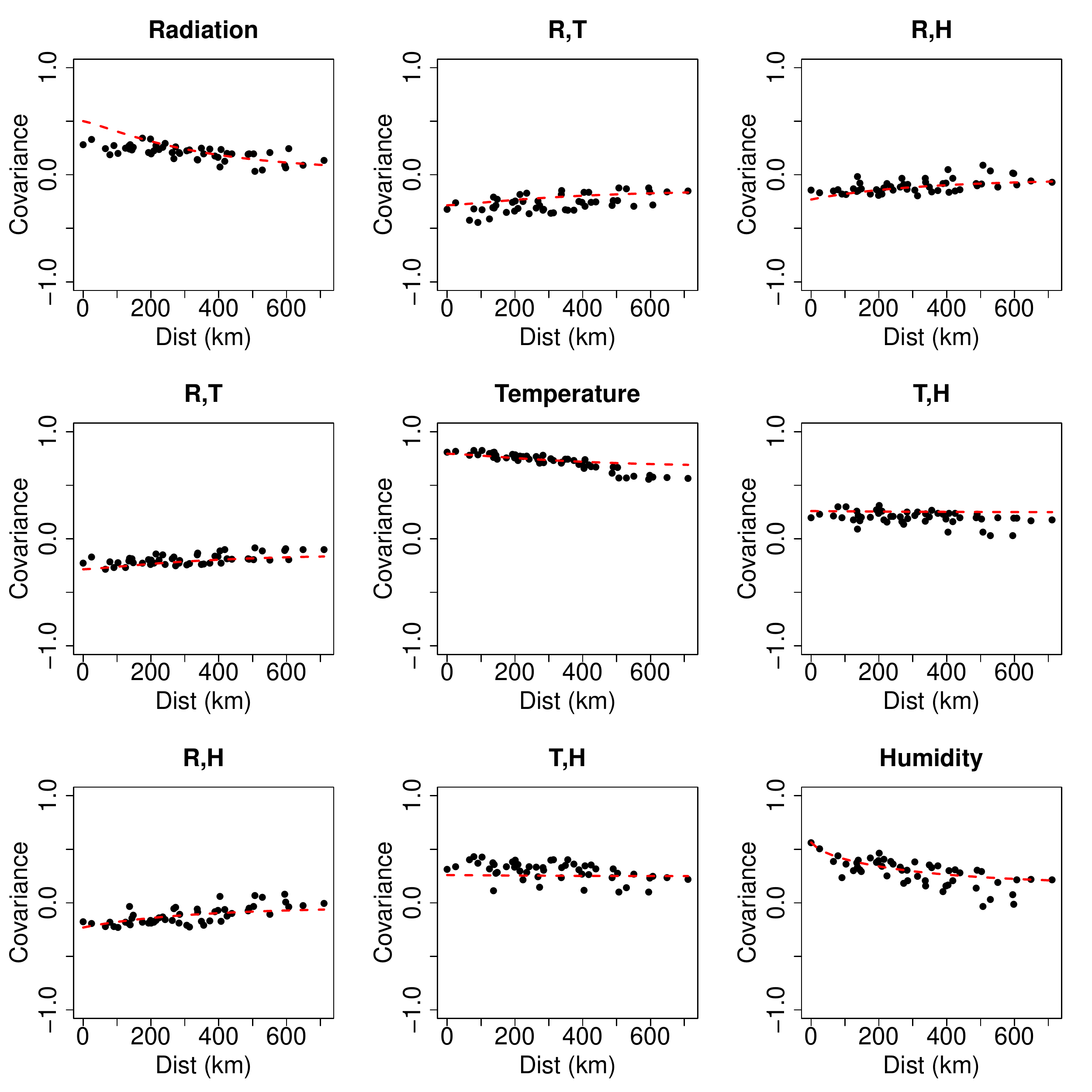}
  \end{center}

      \caption{Spatial direct and cross-covariance functions for R, T and H. Points: empirical values. Solid line: model NS-D with parameters estimated using WPL 
(see Table \ref{tab:estimates}). Left panel: $u=0$. Right panel: $u=1$.\label{fig:covSpatial}}
      
\end{figure}

\begin{figure}

  \begin{center}
    \includegraphics[width=7.5cm]{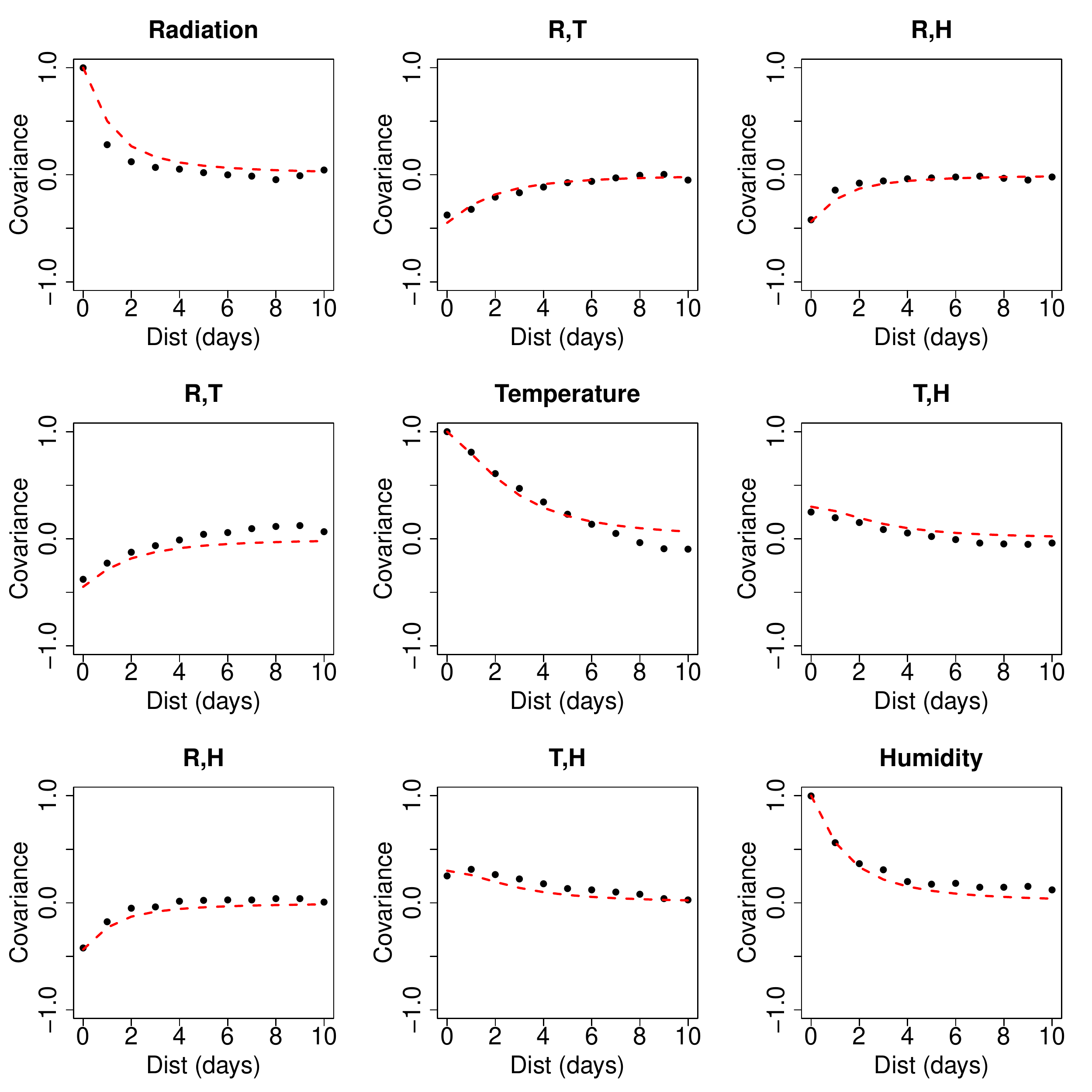} \quad
    \includegraphics[width=7.5cm]{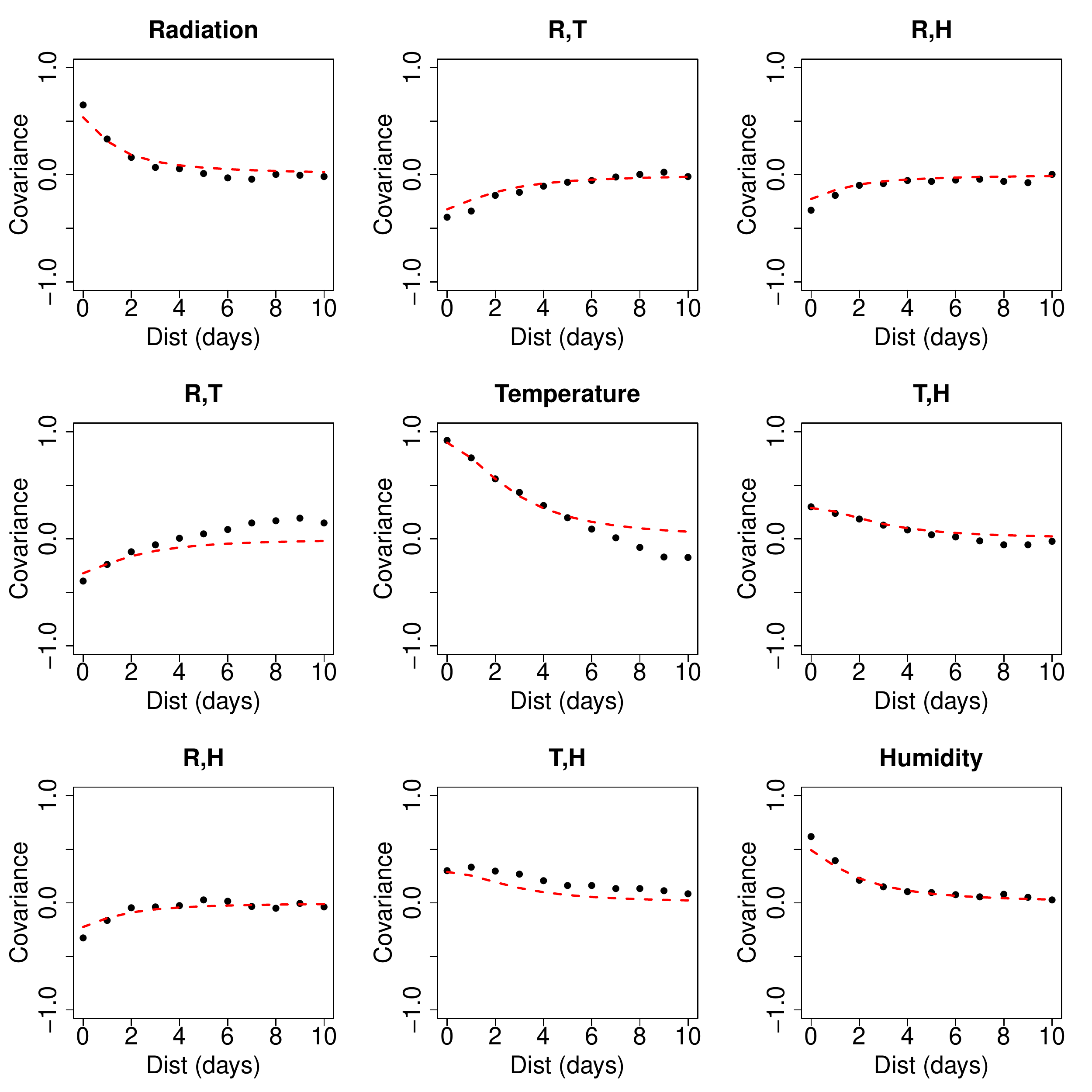}
  \end{center}

      \caption{Temporal direct and cross-covariance functions for R, T and H, as in Figure \ref{fig:covSpatial}. Left panel: ${\bf h} = 0$. Right panel:
$\|{\bf h}\|=199$~km.  \label{fig:covTemporal}}
  
\end{figure}

\begin{figure}

  \begin{center}
    \includegraphics[width=7.5cm]{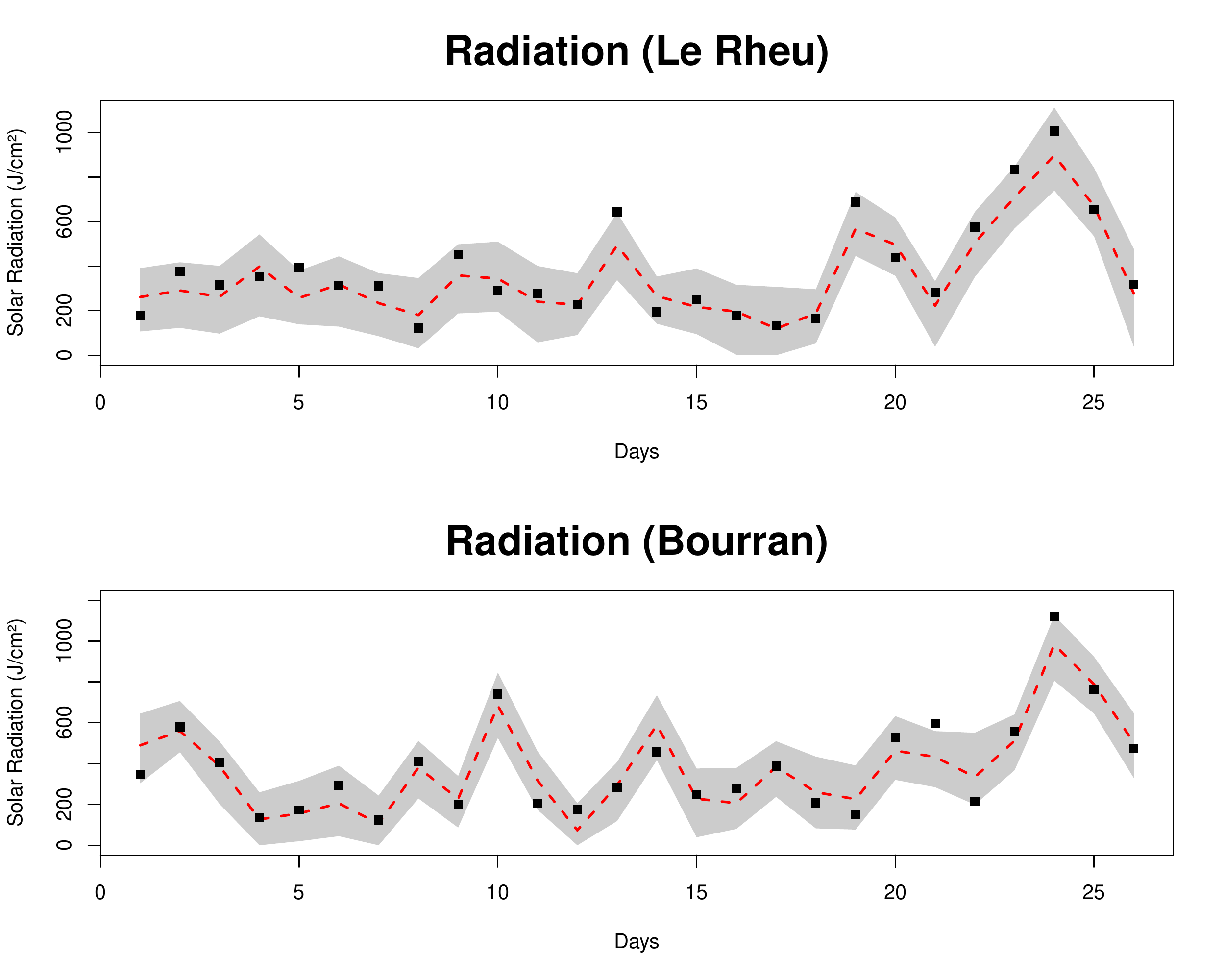} 

    \includegraphics[width=7.5cm]{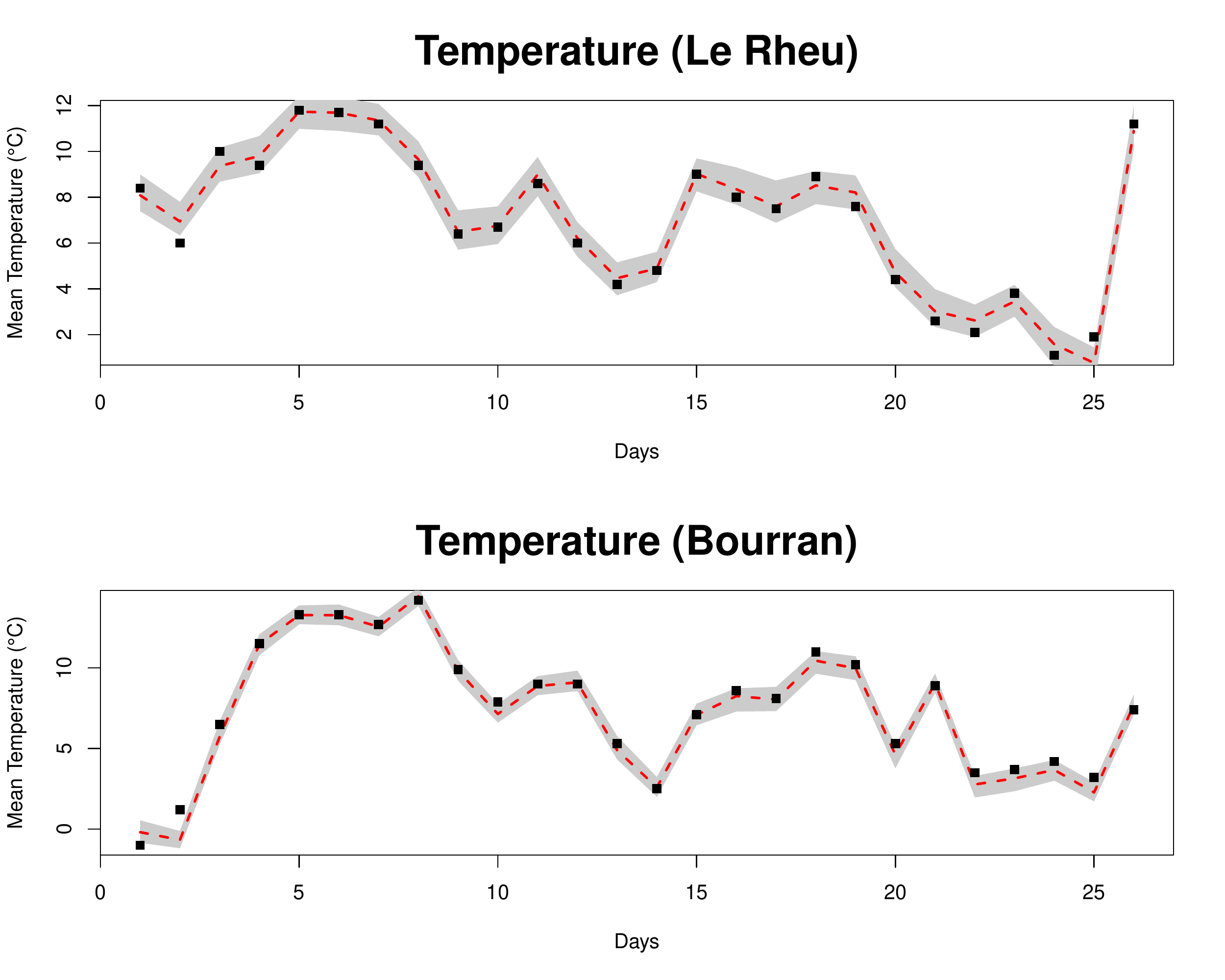}

    \includegraphics[width=7.5cm]{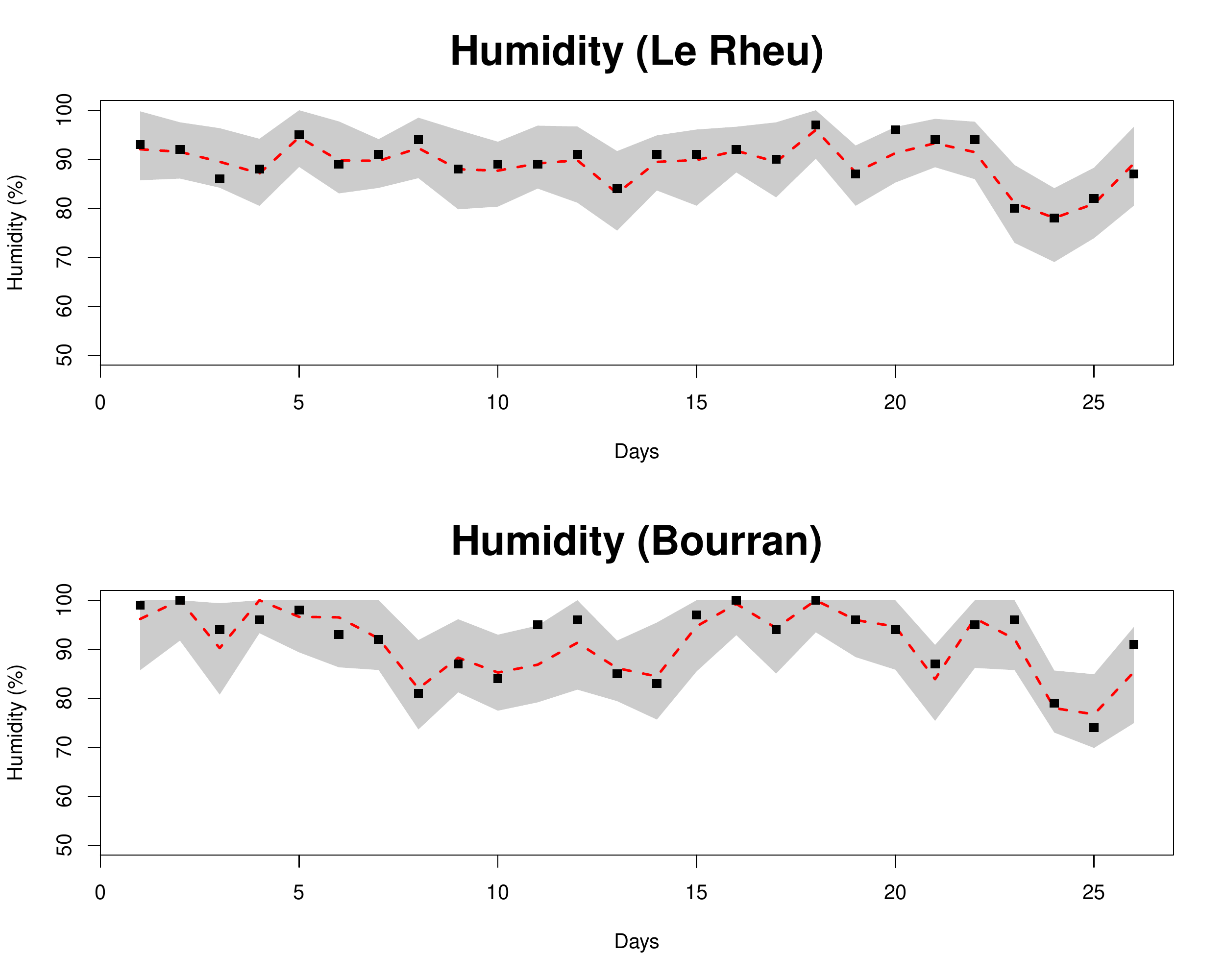}
  \end{center}
 \caption{Prediction of Radiation, Temperature and Humidity at the two validation stations, Le Rheu (Brittany) and Bourran (Aquitaine). Black points: true values. Dotted lines: 
conditional expectation. Shaded area: 90\% envelope interval, computed on 100 simulations.  \label{fig:simuCond} }
 \end{figure}

Regarding the process $\bf W$, we consider the same four space-time models as in the previous section, for which there are at most
15 parameters to be estimated (for model NS-D). There are 8 parameters for the process $\bf X$. Since the data have been 
standardized, we impose $\left[\sigma_i^{(\textbf{X})}\right]^2 + \left[\sigma_i^{(\textbf{W})}\right]^2=1$, $i \in \{{\rm R,T,H} \}$, 
thus leading to a total of at most 20 parameters. Estimates are reported in Table \ref{tab:estimates}. We also
find that the model NS-E performs
worst than S-E. Figures \ref{fig:covSpatial} and \ref{fig:covTemporal} and represent 
the  empirical spatial direct and cross-covariance functions and the model NS-D with estimated parameters using 
WPL. We observe that, in general, the model fits quite well the experimental covariance. 
In particular, the model is able to fit the different 
regularities, as well as the space-time interactions. It is slightly better fitted for the spatial 
dimensions than along time, probably because of much less flexibility in the model along time. Conditional expectation and envelopes of 100 conditional simulations
are shown in Figure  \ref{fig:simuCond} at the two validation locations. The true values are always contained within the envelopes.

\medskip

The analysis of the colocated covariance coefficients between the three variables reveals an interesting pattern. We compute the estimated 
covariance coefficient $\sigma^{(\textbf{X})}_{ij}=\sigma^{(\textbf{X})}_i\sigma^{(\textbf{X})}_j\rho^{(\textbf{X})}_{ij}$, with $i,j \in \{{\rm R,T,H} \}$,
where $\rho^{(\textbf{X})}_{ij}$ is the correlation coefficient as defined in Theorem \ref{theo:GM}. Similar coefficients are computed for 
$\bf W$ and $\bf Z$ and compared to the empirical correlation coefficients in Table  \ref{tab:correlation}. 
The correlation coefficients estimated for $\bf Z$ are very close
to empirical ones.  Note that Radiation is always negatively correlated to Temperature and Humidity, which is typical for winter weather conditions.
More interestingly, we can observe that the (negative) 
correlation between Radiation and Humidity is mostly due to the space-time process $\bf W$, whereas the (positive) correlation between Temperature and 
Humidity is mostly due to temporal process $\bf X$.

\begin{table}   
\begin{tabular}{l|rrrrrrrrrrrr}
\hline\hline
 & $\sigma_{\rm R}^{(\textbf{X})}$ & $\sigma_{\rm T}^{(\textbf{X})}$ & $\sigma_{\rm H}^{(\textbf{X})}$ & $\beta_{\rm RT}^{(\textbf{X})}$ & $\beta_{\rm RH}^{(\textbf{X})}$ & $\beta_{\rm TH}^{(\textbf{X})}$ 
& $\sigma_{\rm R}^{(\textbf{W})}$ & $\sigma_{\rm T}^{(\textbf{W})}$ & $\sigma_{\rm H}^{(\textbf{W})}$ & $\beta_{\rm RT}^{(\textbf{W})}$ & $\beta_{\rm RH}^{(\textbf{W})}$ & $\beta_{\rm TH}^{(\textbf{W})}$ \\
\hline
S-E    & 0.38 & 0.90 & 0.49 & $-0.72$ & $-0.52$ & 0.66 & 0.93 & 0.43 & 0.87 & $-0.46$ & $-0.45$ & $-0.02$ \\
NS-E  & 0.33 & 0.90 & 0.45 & $-0.79$ & $-0.52$ & 0.71 & 0.94 & 0.44 & 0.89 & $-0.47$ & $-0.45$ & $-0.02$ \\
S-D    & 0.20 & 0.85 & 0.43 & $-0.98$ & $-0.56$ & 0.72 & 0.98 & 0.53 & 0.90 & $-0.53$ & $-0.44$ & 0.08 \\
NS-D  & 0.19 & 0.87 & 0.44 & $-0.97$ & $-0.55$ & 0.73 & 0.98 & 0.50 & 0.90 & $-0.59$ & $-0.47$ & 0.06 \\
\hline
\hline
 & $\nu_{\rm R}$ & $\nu_{\rm T}$ & $\nu_{\rm H}$ & $1/r_{\rm R}$ & $1/r_{\rm T}$ & $1/r_{\rm H}$ & $\alpha^{(\textbf{X})}$ & $a^{(\textbf{X})}$ & $\alpha^{(\textbf{W})}$ & $a^{(\textbf{W})}$ & $b$ & -wcl \\
\hline
S-E  & 0.66 & 0.66 & 0.66 & 184 & 184 & 184 & 0.12 & 1.00 & 2.35 & 0.89 & 0 & 631995 \\
NS-E & 0.67 & 0.67 & 0.67 & 188 & 188 & 188 & 0.13 & 0.93 & 2.08 & 1.00 & 0.6 & 631991 \\
S-D  & 0.75 & 0.84 & 0.41 & 211 & 249 & 248 & 0.09 & 1.00 & 0.91 & 0.69 & 0 & 631922 \\
NS-D & 0.64 & 0.84 & 0.30 & 239 & 224 & 322 & 0.12 & 1.00 & 1.06 & 0.75 & 0.6 & {\bf 631845} \\
\hline
\hline
\end{tabular}
\caption{Estimates of the parameters using WPL with $\textbf{d} =(500{\rm km},2{\rm days})$. Unit is kms for distances.\label{tab:estimates}}
\end{table}

\begin{table} 
\centering
\begin{tabular}{c|rrrr}
\hline \hline 
 & RMSE & MAE & CRPS & LogS$_1$ \\
\hline
S-E & 0.433 & 0.306 & 0.222 & 0.440 \\
NS-E & 0.446 & 0.313 & 0.228 & 0.467 \\
S-D & 0.417 & 0.295 & {\bf 0.215} & 0.405 \\
NS-D & {\bf 0.417} & {\bf 0.295} & 0.216 & {\bf 0.397} \\ 
\hline \hline 
\end{tabular}
\caption{RMSE, MAE, CRPS and LogS$_1$ of predicted values at the validation stations, using WPL estimates with $\textbf{d}=(500{\rm km},2{\rm days})$. \label{tab:scores}}

\end{table}

\begin{table} 
\centering
\begin{tabular}{l|ccc}
\hline\hline
 & $\sigma_{\rm RT}$ & $\sigma_{\rm RH}$ & $\sigma_{\rm TH}$ \\
\hline
Process \textbf{X} & $-0.16$ & $-0.05$ & 0.28 \\
Process \textbf{W} & $-0.29$ & $-0.39$ & 0.03 \\
Process \textbf{Z} & $-0.45$ & $-0.44$ & 0.31 \\
Empirical          & $-0.38$ & $-0.42$ & 0.25 \\ 
\hline\hline
\end{tabular}
\caption{Empirical and estimated covariance coefficients  $\sigma_{ij}=\sigma_i\sigma_j\rho_{ij}$, with 
$i,j \in \{{\rm R,T,H} \}$ for the model NS-D.\label{tab:correlation} }
\end{table}

\section{Discussion}

Flexible multivariate space-time models can be built using a mixture approach. We proposed two classes based on the univariate Gneiting class of space-time covariances: 
the Gneiting-Mat\'ern and the Gneiting-Cauchy multivariate space-time models. A simulation study showed that the relatively large number of parameters could be 
accurately estimated by maximizing a weighted pairwise likelihood function. Whenever data are simulated from the most general model, the analysis of the estimation bias and
variance, as well as that of validation scores revealed that ignoring space-time interaction or the multivariate flexibility consistently leads to non optimal scores. 
The relative efficiency of estimates obtained when maximizing  weighted pairwise likelihood with respect to those obtained when maximizing a full likelihood is, 
in general, above 70\%. It thereby confirms that  weighted pairwise likelihood is a reliable alternative to full likelihood when analyzing large datasets.

\medskip

Our approach could be generalized in several ways. First, with similar arguments to those used in Theorems \ref{theo:GM} and \ref{theo:GC}, we could propose a wider range
of flexible  multivariate space-time models, including compactly supported models, and models on manifolds such as spheres. One limit of our models 
is that there is a unique temporal covariance for all variables. By exchanging the role of time and space, one can easily build models that are flexible in time and unique in space.
Building valid multivariate space-time models that are flexible both in time and space is still an open challenge. For the time being, the analysis of models requiring 
different temporal covariances for different variables can be carried out using a sum of models, as in Section \ref{sec:data}. 

\medskip

WPL has been shown to provide accurate estimates for univariate spatial and space-time Gaussian random fields \citep{Bevilacqua2012,Bevilacqua2014}. We showed that it
also provides
accurate estimates for multivariate space-time Gaussian random fields. We found, however, that the space-time separability parameter, $b$, 
was particularly difficult to estimate. The effect of the  separability, which could be formally defined as the ratio between the non-separable space-covariance and 
the separable one, decreases as $\|{\bf h}\| \to 0$ or $u \to 0$. Therefore, the relative efficiency of the estimator of $b$ decreases when the 
window $\textbf{d}=(d_S,d_T)$ defining the pairs in WPL is too small. The optimal size of this
window is thus a balance between two opposite requirements: accurately estimating $b$, while offering speed gain. Our opinion is that one could probably increase 
the efficiency of the estimation of the parameters without sacrificing too much computation speed by proposing new forms of composite likelihoods.

\section*{Appendix}

Our proofs are based on the scale mixture representation (\ref{eq:eq_proof}), for which, according to Lemma 1, we need the following ingredients: 
a relevant univariate covariance function $C_\xi(\textbf{h},u)$ in $\mathbb{R}^k\times\mathbb{R}$ and a non-negative
definite multivariate mixture $\textbf{m}(\xi)=\left[m_{ij}(\xi)\right]_{i,j=1}^p,\xi>0$. The measure $F$ in Lemma 1 will be set 
to be the Lebesgue measure on $(0,\infty)$.

\medskip

The first part of the proof is common to both Theorems. We first verify that the function 
\begin{equation}
C_\xi(\textbf{h},u)=\frac{1}{\psi(u^2)^{k/2}}\exp\left\{-\xi\left(\frac{\|\textbf{h}\|}{\psi(u^2)^{1/2}}\right)^\lambda \right\},
\label{eq:eq_proof_1}
\end{equation}
defined on $\mathbb{R}^k\times\mathbb{R}$ with $0<\lambda\leq2$, is a valid covariance function for any
$\xi > 0$ and for $\psi(t),t\geq 0,$ being a positive function with a completely monotone derivative. Indeed, the mapping
$t \mapsto \exp(-c t^\gamma),\ t\geq 0, \ c>0$ and  $0<\gamma\leq 1$ is completely monotone. Then, setting
$\gamma=1$, $c=\left(\frac{\|\textbf{h}\|}{\psi(u^2)^{1/2}}\right)^\lambda$ and $t=\xi$, it is clear that 
 $C_\xi(\textbf{h},u)$ belongs to the Gneiting class of covariance functions \citep{Gneiting2002} for any $\xi >0$. Also, the mapping
$[0,\infty) \to \mathbb{R}:\ \xi \mapsto C_{\xi} (\textbf{h},u)$ satisfies the hypothesis of Lemma 1, for all fixed $(\textbf{h},u) \in 
\mathbb{R}^k\times\mathbb{R}$. 

\medskip

The rest of the proofs is splitted into two parts corresponding to each model. It consists in finding multivariate 
mixtures $\textbf{m}(\xi)=\left[m_{ij}(\xi)\right]_{i,j=1}^p \in \mathbb{R}^p \times \mathbb{R}^p$ tailored
to either the Mat\'ern or the Cauchy class, that are
symmetric and non-negative definite for all $\xi\geq 0$.

\medskip

\noindent \textbf{Proof of Theorem 1} 

Let $\nu >0$ and $r>0$. From Eq. 3.471.9 in \citet{Gradshteyn2007}, we have
\begin{equation*}
{\cal M}(\textbf{h}; r,\nu)=\int_0^\infty \exp\{-\|\textbf{h}\|^2\xi\}m^{\cal M}(\xi;r,\nu){\rm d}\xi,
\end{equation*}
with
\begin{equation*}
m^{\cal M}(\xi;r,\nu)=\left(\frac{r^2}{4}\right)^\nu \frac{\xi^{-1-\nu}}{\Gamma(\nu)}\exp\left\{-\frac{r^2}{4\xi}\right\}, \quad \xi > 0.
\end{equation*}
Following the univariate mixture representation above, we set 
\begin{equation*}
 \textbf{m}^\mathcal{M}(\xi) = \left[m_{ij}^\mathcal{M} (\xi)\right]_{i,j=1}^p =
\left[\rho_{ij}\left(\frac{r_{ij}^2}{4}\right)^{\nu_{ij}}\frac{\xi^{-1-\nu_{ij}}}{\Gamma(\nu_{ij})}\exp\left\{-\frac{r_{ij}^2}{4\xi}\right\}\right]_{i,j=1}^p, \quad \xi > 0,
\end{equation*}
with 
\begin{eqnarray*}
r_{ij}&= &\{
(r_i^2+r_j^2)/2\}^{1/2},\ \forall i,j=1,\dots,p\text{ and }r_i>0, \ \forall i=1,\dots,p,\\
\nu_{ij}&=&(\nu_i+\nu_j)/2,\ \forall i,j=1,\dots,p,\\
\rho_{ij}&=&\beta_{ij}\frac{\Gamma(\nu_{ij})}{\Gamma(\nu_i)^{1/2}\Gamma(\nu_j)^{1/2}}\frac{r_i^{\nu_i}r_j^{\nu_j}}{r_{ij}^{2\nu_{ij}}}
,\ \forall i,j=1,\dots,p,
\end{eqnarray*}
where $\boldsymbol{\beta} = \left[\beta_{ij}\right]_{i,j=1}^p$ is a correlation matrix. We have, for each $i,j = 1,\dots,p$,
\begin{eqnarray*}
m_{ij}^\mathcal{M}(\xi)&=&\rho_{ij}\left(\frac{r_{ij}^2}{4}\right)^{\nu_{ij}}\frac{\xi^{-1-\nu_{ij}}}{\Gamma(\nu_{ij})}\exp\left\{-\frac{r_{ij}^2}{4\xi} \right\}\\
&=&\beta_{ij}\ \frac{\Gamma(\nu_{ij})}{\Gamma(\nu_i)^{1/2}\Gamma(\nu_j)^{1/2}}\frac{r_i^{\nu_i}r_j^{\nu_j}}{r_{ij}^{2\nu_{ij}}}
\left(\frac{r_{ij}^2}{4}\right)^{\nu_{ij}}\frac{\xi^{-1-\nu_{ij}}}{\Gamma(\nu_{ij})}\exp\left\{-\frac{r_{ij}^2}{4\xi}\right\}\\
&=&\beta_{ij}\ \frac{r_i^{\nu_i}}{2^{\nu_i}\Gamma(\nu_i)^{1/2}}\xi^{-(1+\nu_i)/2}
\exp\left\{-\frac{r_i^2}{8\xi}\right\}\cdot\frac{r_j^{\nu_j}}{2^{\nu_j}\Gamma(\nu_j)^{1/2}}\xi^{-(1+\nu_j)/2}
\exp\left\{-\frac{r_j^2}{8\xi}\right\}\\
&=&\beta_{ij}\ m_i^\mathcal{M}(\xi)^{1/2}m_j^\mathcal{M}(\xi)^{1/2}.
\end{eqnarray*}
Therefore $\textbf{m}^\mathcal{M}(\xi)$ is the Hadamard product of the correlation matrix $\boldsymbol{\beta}$ and the outer product 
$\bar{\textbf{m}}^{\cal M}(\xi)$ by itself, with $\bar{\textbf{m}}^{\cal M}(\xi) = (m_1^{\cal M}(\xi)^{1/2},\dots,m_p^{\cal M}(\xi)^{1/2})^\top$. 
By Schur's Theorem \citep[p. 455]{Horn2012}, the matrix $\textbf{m}^\mathcal{M}(\xi)$ is thus non-negative definite since
it is the Hadamard product of two non-negative definite matrices.

\medskip

Setting $\lambda=2$ in Eq. (\ref{eq:eq_proof_1}) with
$\textbf{m}^{\cal M}(\xi)=\left[m_{ij}^{\cal M}(\xi)\right]_{i,j=1}^p$ as defined above leads thus to the
valid $p$-variate matrix-valued covariance function  $\left[C^{\cal M}_{ij}(\textbf{h},u)\right]_{i,j=1}^p$ on $\mathbb{R}^k\times\mathbb{R}$
by application of Lemma 1:
\begin{eqnarray*}
C^{\cal M}_{ij}(\textbf{h},u)&=&\sigma_i\sigma_j\int_0^\infty \frac{1}{\psi(\|\textbf{u}\|^2)^{k/2}} 
\exp\left\lbrace -\xi\left(\frac{\|\textbf{h}\|}{\psi(u^2)^{1/2}}\right)^2\right\rbrace m^{\cal M}_{ij}(\xi){\rm d}\xi\\
&=&\frac{\sigma_i\sigma_j}{\psi(u^2)^{k/2}}\rho_{ij}{\cal M}\left(\frac{\textbf{h}}{\psi(\|\textbf{u}\|^2)^{1/2}};r_{ij},\nu_{ij}\right)
\end{eqnarray*}
$\square$

\medskip

\noindent \textbf{Proof of Theorem 2}

It is well known that the Cauchy covariance function is the Laplace transform of a Gamma distribution. Therefore, we set
\begin{equation*}
 \textbf{m}^{\cal C}(\xi) = \left[m_{ij}^\mathcal{C} (\xi)\right]_{i,j=1}^p =
\left[\rho_{ij} \frac{1}{r_{ij}^{\nu_{ij}}}
\frac{\xi^{\nu_{ij}-1}}{\Gamma(\nu_{ij})}\exp\left\{-\frac{1}{r_{ij}}\xi\right\}
\right]_{i,j=1}^p, \quad \xi >0,
\end{equation*}
with 
\begin{eqnarray*}
r_{ij}&=&\{
(r_i^{-1}+r_j^{-1})/2\}^{-1},\ \forall i,j=1,\dots,p\text{ and }r_i>0, \ \forall i=1,\dots,p,\\
\nu_{ij}&=&(\nu_i+\nu_j)/2,\ \forall i,j=1,\dots,p,\\
\rho_{ij}&=&\beta_{ij}\frac{\Gamma(\nu_{ij})}{\Gamma(\nu_i)^{1/2}\Gamma(\nu_j)^{1/2}}
\frac{r_{ij}^{\nu_{ij}}}{(r_i^{\nu_i}r_j^{\nu_j})^{1/2}}
,\ \forall i,j=1,\dots,p,
\end{eqnarray*}
where $\boldsymbol{\beta} = \left[\beta_{ij}\right]_{i,j=1}^p$ is a correlation matrix. This multivariate mixture is non-negative definite. Indeed, we have
\begin{eqnarray*}
m_{ij}^\mathcal{C}(\xi)&=&
\rho_{ij} \frac{1}{r_{ij}^{\nu_{ij}}}
\frac{\xi^{\nu_{ij}-1}}{\Gamma(\nu_{ij})}\exp\left\{-\frac{1}{r_{ij}}\xi\right\}\\
&=&\beta_{ij}\ \frac{\Gamma(\nu_{ij})}{\Gamma(\nu_i)^{1/2}\Gamma(\nu_j)^{1/2}}
\frac{r_{ij}^{\nu_{ij}}}{(r_i^{\nu_i}r_j^{\nu_j})^{1/2}}
\frac{1}{r_{ij}^{\nu_{ij}}}\frac{\xi^{\nu_{ij}-1}}{\Gamma(\nu_{ij})}\exp\left\{-\frac{1}{r_{ij}}\xi\right\}\\
&=&\beta_{ij}\ \frac{\xi^{\frac{\nu_i-1}{2}}}{(\Gamma(\nu_i)r_i^{\nu_i})^{1/2}}
\exp\left\{-\frac{1}{2r_i}\xi\right\}\cdot
\frac{\xi^{\frac{\nu_j-1}{2}}}{(\Gamma(\nu_j)r_j^{\nu_j})^{1/2}}
\exp\left\{-\frac{1}{2r_j}\xi\right\}\\
&=&\beta_{ij}\ m_i^\mathcal{C}(\xi)^{1/2}m_j^\mathcal{C}(\xi)^{1/2}.
\end{eqnarray*}
The same arguments as in Theorem 1 are used to conclude that  $\textbf{m}^{\cal C}(\xi)$ is non-negative definite. 
The last step of the proof consists in applying Lemma 1 with $\textbf{m}^{\cal C}(\xi)=\left[m_{ij}^{\cal C}(\xi)\right]_{i,j=1}^p$ as defined above
and by setting $\lambda=2$ in Eq. (\ref{eq:eq_proof_1}).
The result is the valid $p$-variate matrix-valued covariance function on $\mathbb{R}^k\times\mathbb{R}$

\begin{eqnarray*}
C^{\cal C}_{ij}(\textbf{h},u)&=&\sigma_i\sigma_j\int_0^\infty \frac{1}{\psi(u^2)^{k/2}} 
\exp\left\lbrace -\xi\left(\frac{\|\textbf{h}\|}{\psi(u^2)^{1/2}}\right)^\lambda\right\rbrace m^{\cal C}_{ij}(\xi){\rm d}\xi\\
&=&\frac{\sigma_i\sigma_j}{\psi(u^2)^{k/2}}\rho_{ij}
\int_0^\infty  
\exp\left\lbrace -\xi\left(\frac{\|\textbf{h}\|}{\psi(u^2)^{1/2}}\right)^\lambda\right\rbrace
\frac{1}{r_{ij}^{\nu_{ij}}}
\frac{\xi^{\nu_{ij}-1}}{\Gamma(\nu_{ij})}\exp\left\{-\frac{1}{r_{ij}}\xi\right\}{\rm d}\xi
\\
&=&\frac{\sigma_i\sigma_j}{\psi(u^2)^{k/2}}\rho_{ij}
\left\lbrace 1+r_{ij}\left(\frac{\|\textbf{h}\|}{\psi(u^2)^{1/2}}\right)^\lambda\right\rbrace^{-\nu_{ij}}
\\
& &\ \ \ \times\int_0^\infty  
\left\lbrace \frac{1}{r_{ij}}+\left(\frac{\|\textbf{h}\|}{\psi(u^2)^{1/2}}\right)^\lambda\right\rbrace^{\nu_{ij}}
\frac{\xi^{\nu_{ij}-1}}{\Gamma(\nu_{ij})}
\exp\left\lbrace -\xi\left[\frac{1}{r_{ij}}+\left(\frac{\|\textbf{h}\|}{\psi(u^2)^{1/2}}\right)^\lambda\right]\right\rbrace
{\rm d}\xi
\\
&=&\frac{\sigma_i\sigma_j}{\psi(u^2)^{k/2}}\rho_{ij}\ {\cal C}\left(\frac{\textbf{h}}{\psi(u^2)^{1/2}}; r_{ij},\nu_{ii},\lambda\right) \square
\end{eqnarray*}

\section*{Acknowledgments} Two anonymous referees are gratefully acknowledged.
The authors wish to thank Carlo Gaetan from the Ca' Foscari University in Venezia (Italy)
for fruitful discussions and Liliane Bel from AgroParisTech in Paris (France) for her very careful reading of earlier 
versions of this manuscript. This work was supported by the metaprogramme Adaptation of Agriculture and Forests to Climate 
Change (AAFCC) of the French National Institute for Agricultural Research (INRA).
Emilio Porcu has been supported by Proyecto Fondecyt n. 1130647 from Chilean Ministry of Education.

\bibliographystyle{apalike}

\begin{thebibliography}{}

\bibitem[Abramowitz and Stegun, 1972]{Abramowitz1972}
Abramowitz, M. and Stegun, I.~A. (1972).
\newblock {\em Handbook of mathematical functions: with formulas, graphs, and
  mathematical tables}.
\newblock Number~55. Courier Dover Publications.

\bibitem[Apanasovich and Genton, 2010]{Apanasovich2010}
Apanasovich, T.~V. and Genton, M.~G. (2010).
\newblock Cross-covariance functions for multivariate random fields based on
  latent dimensions.
\newblock {\em Biometrika}, 97(1):15--30.

\bibitem[Apanasovich et~al., 2012]{Apanasovich2012}
Apanasovich, T.~V., Genton, M.~G., and Sun, Y. (2012).
\newblock A valid {M}at{\'e}rn class of cross-covariance functions for
  multivariate random fields with any number of components.
\newblock {\em Journal of the American Statistical Association},
  107(497):180--193.

\bibitem[Bevilacqua and Gaetan, 2015]{Bevilacqua2014}
Bevilacqua, M. and Gaetan, C. (2015).
\newblock Comparing composite likelihood methods based on pairs for spatial
  {G}aussian random fields.
\newblock {\em Statistics and Computing}, 25(5):877--892.

\bibitem[Bevilacqua et~al., 2012]{Bevilacqua2012}
Bevilacqua, M., Gaetan, C., Mateu, J., and Porcu, E. (2012).
\newblock Estimating space and space-time covariance functions for large data
  sets: a weighted composite likelihood approach.
\newblock {\em Journal of the American Statistical Association},
  107(497):268--280.

\bibitem[Castruccio et~al., 2015]{Castruccio2014}
Castruccio, S., Huser, R., and Genton, M. (2015).
\newblock High-order composite likelihood inference for max-stable
  distributions and processes.
\newblock arXiv:1411.0086v3.

\bibitem[Cressie, 1993]{Cressie1993}
Cressie, N. (1993).
\newblock {\em Statistics for Spatial Data: Wiley Series in Probability and
  Statistics}.
\newblock Wiley-Interscience New York.

\bibitem[Cressie and Wikle, 2011]{Cressie2011}
Cressie, N. and Wikle, C.~K. (2011).
\newblock {\em Statistics for spatio-temporal data}.
\newblock John Wiley \& Sons.

\bibitem[Daley et~al., 2014]{Daley2014}
Daley, D., Porcu, E., and Bevlicqua, M. (2014).
\newblock Classes of compactly supported covariance functions for multivariate
  random fields.
\newblock {\em Stochastic Environmental Research and Risk Assessment},
  29(4):1249--1263.

\bibitem[De~Iaco et~al., 2013]{DeIaco2013}
De~Iaco, S., Myers, D., Palma, M., and Posa (2013).
\newblock Using simultaneous diagonalization to identify a space-time linear
  coregionalization model.
\newblock {\em Mathematical Geosciences}, 45(1):69--86.

\bibitem[Gelfand and Banerjee, 2010]{GelfandBanerjee2010}
Gelfand, A.~E. and Banerjee, S. (2010).
\newblock Multivariate spatial process models.
\newblock In Gelfand, A.~E., Diggle, P., Fuentes, M., and Guttorp, P., editors,
  {\em Handbook of Spatial Statistics}, pages 495--515. Chapman \& Hall/CRC.

\bibitem[Genton and Kleiber, 2015]{GentonKleiber2014}
Genton, M.~G. and Kleiber, W. (2015).
\newblock Cross-covariance functions for multivariate geostatistics.
\newblock {\em Statistical Science}, 30(2):147--163.

\bibitem[Gneiting, 2002]{Gneiting2002}
Gneiting, T. (2002).
\newblock Nonseparable, stationary covariance functions for space--time data.
\newblock {\em Journal of the American Statistical Association},
  97(458):590--600.

\bibitem[Gneiting et~al., 2006]{Gneiting2007}
Gneiting, T., Genton, M., and Guttorp, P. (2006).
\newblock Geostatistical space-time models, stationarity, separability and full
  symmetry.
\newblock In Fintenst\"adt, B., Held, L., and Isham, V., editors, {\em
  Statistical Methods for Spatio-Temporal Systems}, pages 151--175. Chapman \&
  Hall/CRC.

\bibitem[Gneiting et~al., 2010]{Gneiting2010}
Gneiting, T., Kleiber, W., and Schlather, M. (2010).
\newblock Mat{\'e}rn cross-covariance functions for multivariate random fields.
\newblock {\em Journal of the American Statistical Association},
  105(491):1167--1177.

\bibitem[Gneiting and Raftery, 2007]{GneitingRaftery2007}
Gneiting, T. and Raftery, A.~E. (2007).
\newblock Strictly proper scoring rules, prediction, and estimation.
\newblock {\em Journal of the American Statistical Association},
  102(477):359--378.

\bibitem[Gneiting and Schlather, 2004]{GneitingSchlather2004}
Gneiting, T. and Schlather, M. (2004).
\newblock Stochastic models that separate fractal dimension and the {H}urst
  effect.
\newblock {\em SIAM Review}, 46(2):269--282.

\bibitem[Goulard and Voltz, 1992]{GoulardVoltz1992}
Goulard, M. and Voltz, M. (1992).
\newblock Linear coregionalization model: Tools for estimation and choice of
  cross-variogram matrix.
\newblock {\em Mathematical Geology}, 24(3):269--286.

\bibitem[Gradshteyn and Ryzhik, 2007]{Gradshteyn2007}
Gradshteyn, I. and Ryzhik, I. (2007).
\newblock {\em Table of integrals, series, and products, Seventh edition}.
\newblock Academic Press.

\bibitem[Horn and Johnson, 2012]{Horn2012}
Horn, R.~A. and Johnson, C.~R. (2012).
\newblock {\em Matrix analysis}.
\newblock Cambridge University Press.

\bibitem[Li et~al., 2008]{Li2008}
Li, B., Genton, M.~G., and Sherman, M. (2008).
\newblock Testing the covariance structure of multivariate random fields.
\newblock {\em Biometrika}, 95(4):813--829.

\bibitem[Lindsay, 1988]{Lindsay1988}
Lindsay, B.~G. (1988).
\newblock Composite likelihood methods.
\newblock {\em Contemporary Mathematics}, 80(1):221--239.

\bibitem[Mat{\'e}rn, 1986]{Matern1986}
Mat{\'e}rn, B. (1986).
\newblock {\em Spatial Variation}.
\newblock Lecture Notes in Statistics, 36, Springer.

\bibitem[Porcu and Schilling, 2011]{Porcu2011}
Porcu, E. and Schilling, R.~L. (2011).
\newblock From {S}choenberg to {P}ick--{N}evanlinna: Toward a complete picture
  of the variogram class.
\newblock {\em Bernoulli}, 17(1):441--455.

\bibitem[Porcu and Zastavnyi, 2011]{PorcuZastavnyi2011}
Porcu, E. and Zastavnyi, V. (2011).
\newblock Characterization theorems for some classes of covariance functions
  associated to vector valued random fields.
\newblock {\em Journal of Multivariate Analysis}, 102:1293--1301.

\bibitem[Reisert and Burkhardt, 2007]{Reisert2007}
Reisert, M. and Burkhardt, H. (2007).
\newblock Learning equivariant functions with matrix valued kernels.
\newblock {\em The Journal of Machine Learning Research}, 8:385--408.

\bibitem[Schlather, 2010]{Schlather2010}
Schlather, M. (2010).
\newblock Some covariance models based on normal scale mixtures.
\newblock {\em Bernoulli}, 16(3):780--797.

\bibitem[Schoenberg, 1938]{Schoenberg1938}
Schoenberg, I.~J. (1938).
\newblock Metric spaces and completely monotone functions.
\newblock {\em Annals of Mathematics}, 39(4):811--841.

\bibitem[Varin et~al., 2011]{Varin2011}
Varin, C., Reid, N., and Firth, D. (2011).
\newblock An overview of composite likelihood methods.
\newblock {\em Statistica Sinica}, 21(1):5--42.

\bibitem[Wackernagel, 2003]{Wackernagel2003}
Wackernagel, H. (2003).
\newblock {\em Multivariate geostatistics}.
\newblock Springer, Berlin.

\bibitem[Zhang, 2004]{Zhang2004}
Zhang, H. (2004).
\newblock Inconsistent estimation and asymptotically equal interpolations in
  model-based geostatistics.
\newblock {\em Journal of the American Statistical Association},
  99(465):250--261.

\end{thebibliography}

\end{document}